\documentclass[amsmath,superscriptaddress,showpacs,showkeys,twocolumn,prl]{revtex4-1}
\usepackage{amssymb,amsmath}   
\usepackage[dvips]{graphicx}   
\usepackage{verbatim}   
\usepackage{color}      
\usepackage{subfigure}  
\usepackage{hyperref}   
\usepackage{gensymb}
\usepackage{epstopdf}
\usepackage{natbib}
\usepackage{braket}
\usepackage{enumerate}

\begin{document}

\title{Dynamical Decoupling of Quantum Two-Level Systems by Coherent Multiple Landau-Zener Transitions}
	
\author{Shlomi Matityahu}\email{matityas@post.bgu.ac.il}
\affiliation{Department of Physics, Ben-Gurion University of the Negev, Beer Sheva 84105, Israel}
\affiliation{Department of Physics, Nuclear Research Centre-Negev, Beer-Sheva 84190, Israel}
\affiliation{Institute of Nanotechnology, Karlsruhe Institute of Technology, D-76344 Eggenstein-Leopoldshafen, Germany}
\author{Hartmut Schmidt}
\affiliation{Physikalisches Institut, Karlsruhe Institute of Technology (KIT), 76131 Karlsruhe, Germany}
\author{Alexander Bilmes}
\affiliation{Physikalisches Institut, Karlsruhe Institute of Technology (KIT), 76131 Karlsruhe, Germany}
\author{Alexander Shnirman}
\affiliation{Institute of Nanotechnology, Karlsruhe Institute of Technology, D-76344 Eggenstein-Leopoldshafen, Germany}
\affiliation{Institut f\"ur Theorie der Kondensierten Materie, KIT, 76131 Karlsruhe, Germany}
\author{Georg Weiss}
\affiliation{Physikalisches Institut, Karlsruhe Institute of Technology (KIT), 76131 Karlsruhe, Germany}
\author{Alexey V. Ustinov}
\affiliation{Physikalisches Institut, Karlsruhe Institute of Technology (KIT), 76131 Karlsruhe, Germany}
\affiliation{Russian Quantum Center, National University of Science and Technology MISIS, Moscow 119049, Russia}
\author{Moshe Schechter}
\affiliation{Department of Physics, Ben-Gurion University of the Negev, Beer Sheva 84105, Israel}
\author{J\"urgen Lisenfeld}
\affiliation{Physikalisches Institut, Karlsruhe Institute of Technology (KIT), 76131 Karlsruhe, Germany}
	
\date{\today}
\begin{abstract}
Increasing and stabilizing the coherence of superconducting quantum circuits and resonators is of utmost importance for various technologies ranging from quantum information processors to highly sensitive detectors of low-temperature radiation in astrophysics. A major source of noise in such devices is a bath of quantum two-level systems (TLSs) with broad distribution of energies, existing in disordered dielectrics and on surfaces. Here we study the dielectric loss of superconducting resonators in the presence of a periodic electric bias field, which sweeps near-resonant TLSs in and out of resonance with the resonator, resulting in a periodic pattern of Landau-Zener transitions. We show that at high sweep rates compared to the TLS relaxation rate, the coherent evolution of the TLS over multiple transitions yields a significant reduction in the dielectric loss relative to the intrinsic value. This behavior is observed both in the classical high-power regime and in the quantum single-photon regime, possibly suggesting a viable technique to dynamically decouple TLSs from a qubit. 
\end{abstract}
	
\keywords{} \maketitle
\section{Introduction} \label{Introduction} 
Superconducting quantum devices are nowadays at the heart of many physical platforms exploring both foundations and applications of quantum mechanics. In particular, superconducting quantum circuits~\cite{WG17} are one of the prime contenders for the realization of a quantum computer~\cite{NC18,OJS17}, and superconducting microwave resonators are of great interest for photon detection in astronomy applications~\cite{DP03,ZJ12}. The coupling of superconducting qubits to resonators provides exciting prospects for studying quantum optics and atomic physics in an engineerable architecture with strong nonlinearities and interactions~\cite{WA04,YJQ11,GX17}.

Originally postulated in the 1970's to explain the low-temperature properties of amorphous solids~\cite{PWA72,AHV72}, tunneling two-level systems (TLSs) have attracted a lot of renewed interest in the field of superconducting quantum devices, where such defects residing in the amorphous oxides of the microfabricated circuits form a major energy relaxation and decoherence channel~\cite{MC18}. Since TLSs couple both to strain and electric fields, those that are in resonance with a device electromagnetic mode efficiently dissipate energy into phonon~\cite{RYJ19} and BCS quasiparticle~\cite{BA17} excitations, giving rise to dielectric loss in superconducting microwave resonators and energy relaxation in superconducting qubits. Moreover, due to mutual TLS-TLS interactions~\cite{LJ15}, the thermal fluctuations of low-frequency TLSs give rise to fluctuations of high-frequency resonant TLSs --- a phenomenon known as spectral diffusion, which causes time-dependent fluctuations of the device's electromagnetic environment~\cite{MC15,MS18,KPV18,NC13,BJ14,FL15,BAL15,SS19,BJ19,ECT18}. Improving and stabilizing the coherence properties of superconducting devices is crucial for the realization of a scalable quantum computer~\cite{NC18,OJS17}.

In the standard tunneling model~\cite{PWA72,AHV72}, each TLS is described by the Hamiltonian
\begin{align}
\label{eq:TLS_Hamiltonian}&
\mathcal{H}=\frac{1}{2}\left(\Delta\sigma^{}_{z}+\Delta^{}_{0}\sigma^{}_{x}\right)+\bigg(\sum^{}_{\alpha,\beta}\gamma^{}_{\alpha\beta}\varepsilon^{}_{\alpha\beta}-\mathbf{p}\cdot\mathbf{E}\bigg)\sigma^{}_{z},
\end{align}
where $\sigma^{}_{x}$ and $\sigma^{}_{z}$ are the Pauli matrices, $\Delta$ and $\Delta^{}_{0}$ are the bias and tunneling energies of the unperturbed TLS, and $\gamma^{}_{\alpha\beta}\equiv(1/2)\partial\Delta/\partial\varepsilon^{}_{\alpha\beta}$, $\mathbf{p}\equiv(1/2)\partial\Delta/\partial\mathbf{E}$ are the elastic quadrupole and electric dipole moments of the TLS, which couple to the strain and electric fields $\varepsilon^{}_{\alpha\beta}$ and $\mathbf{E}$. The distribution of $\Delta$ and $\Delta^{}_{0}$ is quite universal and has the form $f(\Delta,\Delta^{}_{0})=P^{}_{0}/\Delta^{}_{0}$, with $P^{}_{0}$ being a material dependent constant.

For strongly driven superconducting microwave resonators at low temperatures, $k^{}_{\mathrm{B}}T\ll\hbar\omega$, interaction of the resonator electric field $\mathbf{E}^{}_{\mathrm{res}}(t)=\mathbf{E}^{}_{\mathrm{ac}}\cos(\omega t)$ with resonant TLSs leads to the well-known expression for the dielectric loss tangent (inverse quality factor)~\cite{VSM77}, $\tan\delta=\tan\delta^{}_{0}/\sqrt{1+\Omega^{2}_{\mathrm{R0}}T^{}_{1}T^{}_{2}}$. Here $\tan\delta^{}_{0}=\pi P^{}_{0}p^{2}\tanh(\hbar\omega/2k^{}_{\mathrm{B}}T)/(3\epsilon)$ is the intrinsic loss tangent in the low-power limit, with $p=|\mathbf{p}|$ the absolute magnitude of the dipole moment and $\epsilon$ the dielectric constant~\cite{Comment1}, $\Omega^{}_{\mathrm{R}0}=pE^{}_{\mathrm{ac}}/\hbar$ is the TLS (maximum) Rabi frequency (see below) and $T^{}_{1}$, $T^{}_{2}$ are characteristic TLS relaxation and decoherence times. This power dependence arises from saturation of individual TLSs. Unfortunately, using this saturation effect to improve the coherence times of superconducting qubits is impractical, as unwanted qubit excitations are caused either by the applied strong resonant field or by excited TLSs via the qubit-TLS interaction.

Recently, the dielectric loss of superconducting resonators was studied in the presence of a periodic bias field $\mathbf{E}^{}_{\mathrm{bias}}(t)$, which slowly changes the bias energy of TLSs at a rate $v^{}_{0}=2p\dot{E}^{}_{\mathrm{bias}}$, and sweeps them through resonance with the resonator~\cite{KMS14,BAL13}. The dynamics of each transition is of the Landau-Zener (LZ) type~\cite{LLD32,ZC32,SECG32}, with a non-adiabatic transition probability
\begin{align}
\label{eq:LZ_probability}&
P=e^{-1/\xi},
\end{align}
where $\xi=2|v^{}_{0}|/(\pi\hbar\Omega^{2}_{\mathrm{R}0})$ is a dimensionless parameter. At slow sweep rates $|v^{}_{0}|\ll\hbar\Omega^{}_{\mathrm{R}0}\Gamma^{}_{1}$, the transition time for a single LZ transition, $t^{}_{\mathrm{LZ}}=\hbar\Omega^{}_{\mathrm{R}0}/|v^{}_{0}|$, is longer than the TLS relaxation time $T^{}_{1}=\Gamma^{-1}_{1}$; the LZ transitions are irrelevant, and the loss tangent is almost independent of the sweep rate and given by the non-linear saturation discussed above. In terms of $\xi$, this regime can be expressed as $\xi\ll\xi^{}_{1}$, where $\xi^{}_{1}\equiv2\Gamma^{}_{1}/(\pi\Omega^{}_{\mathrm{R}0})$. For $\hbar\Omega^{}_{\mathrm{R}0}\Gamma^{}_{1}\ll|v^{}_{0}|\ll\hbar\Omega^{2}_{\mathrm{R}0}$ (equivalently, $\Omega^{-1}_{\mathrm{R}0}\ll t^{}_{\mathrm{LZ}}\ll T^{}_{1}$ or $\xi^{}_{1}\ll\xi\ll 1$), each LZ transition is coherent and adiabatic, with photon absorption probability $1-P\approx 1$, meaning that each TLS swept through resonance dissipates one photon. As the number of TLSs swept through resonance is proportional to $|v^{}_{0}|$, the loss in this regime increases linearly with $|v^{}_{0}|$. In the regime $|v^{}_{0}|\gg\hbar\Omega^{2}_{\mathrm{R}0}$ ($\xi\gg 1$) each transition becomes non-adiabatic, with photon absorption probability $1-P\propto 1/v^{}_{0}$, leading to a universal constant loss tangent independent of the resonator field~\cite{BAL13,KMS14}. This universal constant loss equals the low-power limit $\tan\delta^{}_{0}$, a consequence of a short transition time $t^{}_{\mathrm{LZ}}$ compared to the Rabi oscillation period $\Omega^{-1}_{\mathrm{R}0}$, such that during resonant passages TLSs are not saturated by the resonator ac field.

A crucial assumption of the results described above is the long period of the bias field, $T^{}_{\mathrm{sw}}$, compared to the relaxation time $T^{}_{1}$. In this regime, TLSs relax after each transition, and two subsequent transitions are independent. Here, we explore a regime of shorter periods, $T^{}_{\mathrm{sw}}<T^{}_{1}$, where the coherent evolution during several LZ transitions has to be considered~\cite{SSN10,OWD05,OWD09,SM06,WCM10,IA08,LMD09}. We show theoretically and experimentally that due to interference effects the resonator loss decreases in this regime. This reduction relative to the intrinsic loss is significant, and the loss reaches a value which may be, in principle, even lower than at zero sweep rate. In contrast to the saturation limit at zero sweep rate discussed above, the low loss in the high sweep rate regime $T^{-1}_{\mathrm{sw}}\gg\Gamma^{}_{1}$ is a consequence of a reduced photon absorption probability due to destructive interference between many LZ transitions. Moreover, whereas saturation of photon absorption is obtained by strong resonant driving for $\Omega^{}_{\mathrm{R}0}\gg\Gamma^{}_{1}$, the reduction of the loss in the regime $T^{-1}_{\mathrm{sw}}\gg\Gamma^{}_{1}$ is achieved by application of time-dependent bias fields with frequency $T^{-1}_{\mathrm{sw}}$ much lower than the resonance frequency $\omega/(2\pi)$. We also discuss the single-photon regime, and show experimental evidence for the applicability of the theory in this regime. Since the physics of the single-photon regime corresponds to that of a qubit coupled to a resonant TLS, the results suggest a technique to effectively decouple near-resonant TLSs from a qubit without affecting the qubit state.

\section{Supression of TLS dielectric loss in microwave resonators} \label{Resonators}
\subsection{Theory} \label{Theory}
\begin{figure}[htb!]
\includegraphics[width=0.9\columnwidth,height=5.5cm]{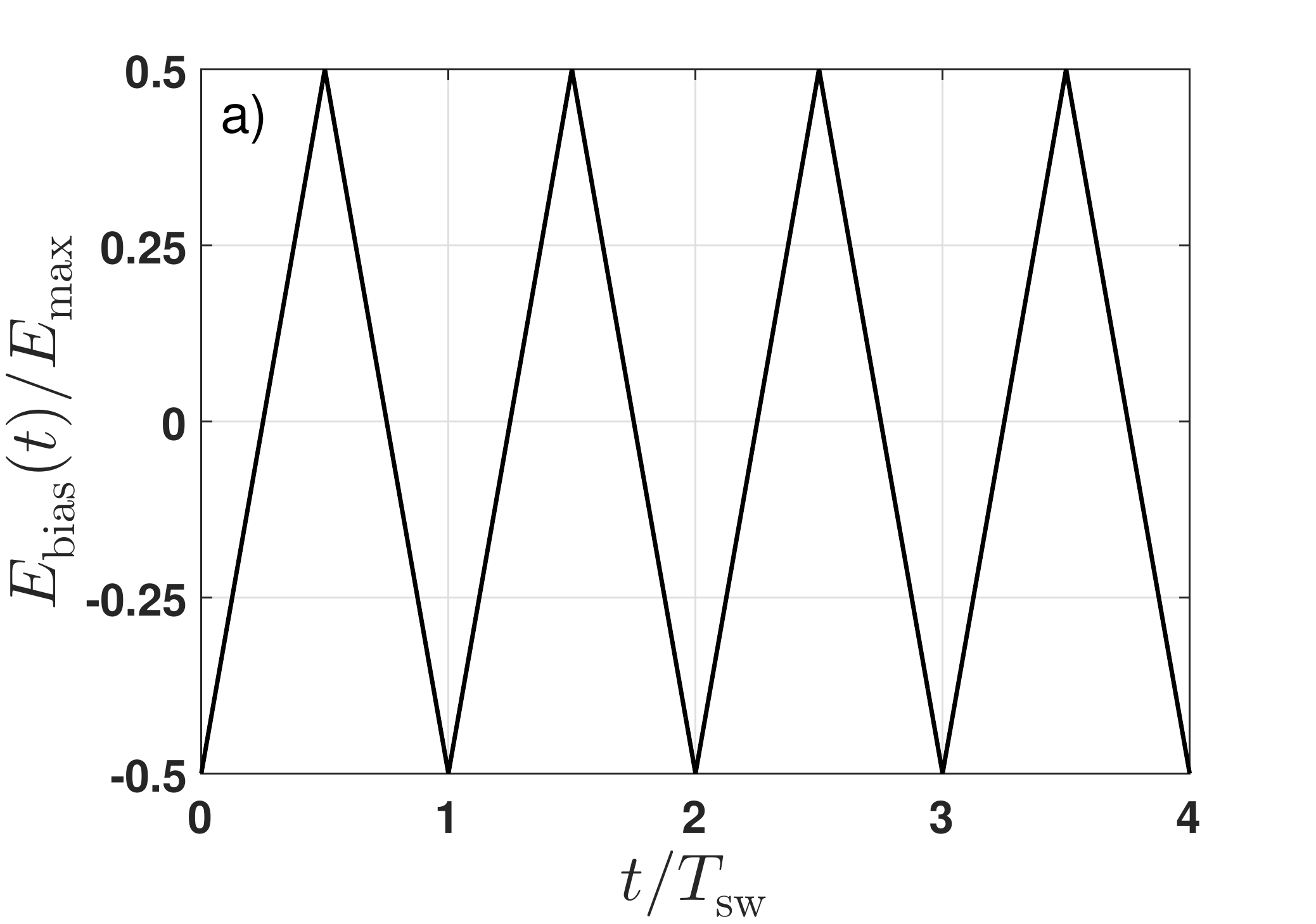}
\includegraphics[width=0.9\columnwidth,height=5.5cm]{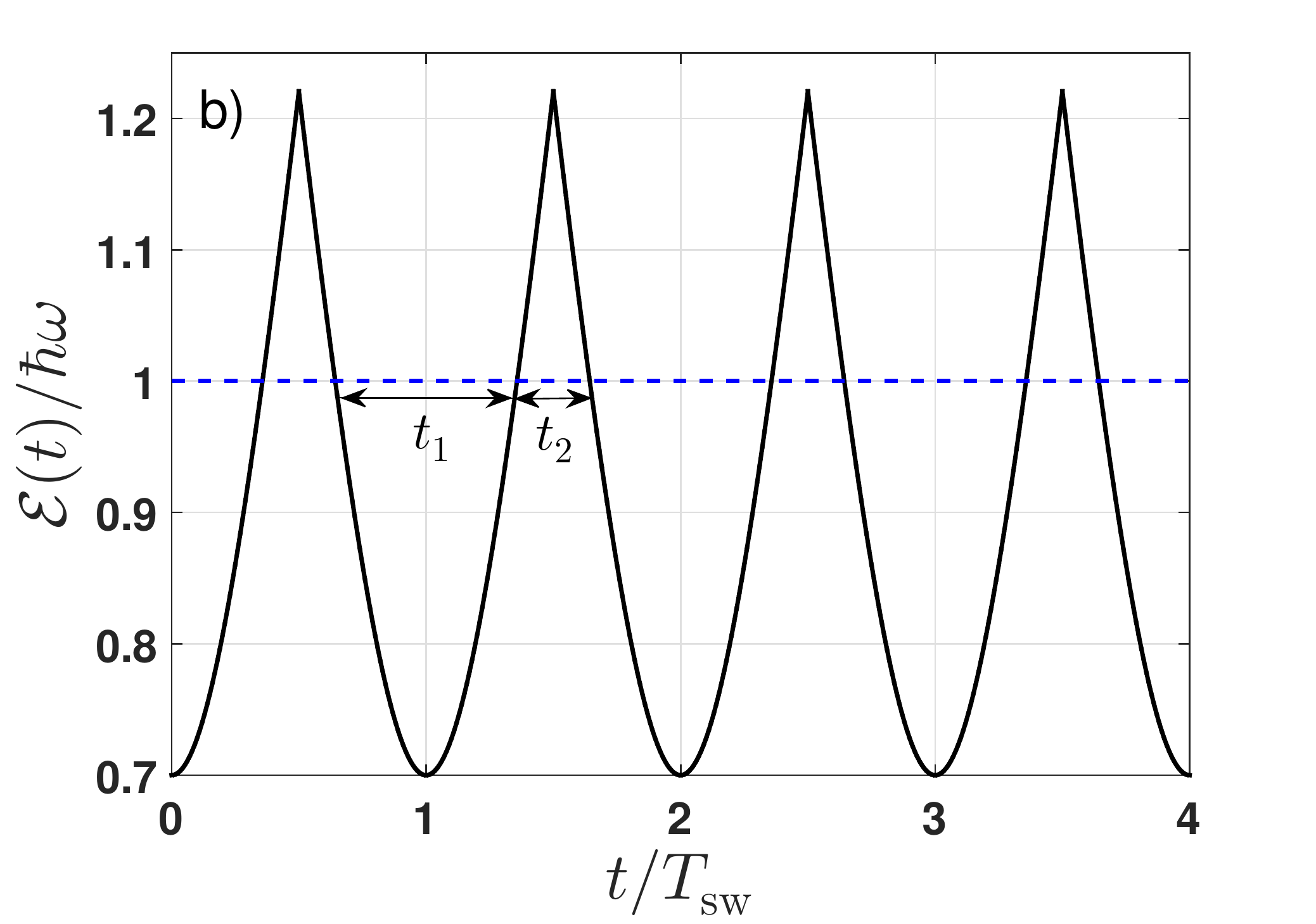}
\caption{a) A triangular wave bias field $E^{}_{\mathrm{bias}}(t)$ and b) the corresponding energy splitting $\mathcal{E}(t)=\sqrt{\Delta^{2}(t)+\Delta^{2}_{0}}$ of a TLS with bias energy $\Delta(t)=0.5+2pE^{}_{\mathrm{bias}}(t)$, tunneling energy $\Delta^{}_{0}=0.7$, and $pE^{}_{\mathrm{max}}=0.5$ (energies are in units of $\hbar\omega$). The intersections of $\mathcal{E}(t)$ with the dashed line correspond to times where the TLS is swept through resonance with the resonator, the dynamics of each resonance is of the LZ type, with the Hamiltonian~(\ref{eq:LZ_Hamiltonian}). In each period of the bias field, the time intervals $t^{}_{1}$ and $t^{}_{2}$ correspond to free propagation between subsequent LZ transitions, with $t^{}_{1}+t^{}_{2}=T^{}_{\mathrm{sw}}$.} 
\label{fig1}
\end{figure}

We consider an arbitrary TLS out of the ensemble of TLSs, described by the Hamiltonian~(\ref{eq:TLS_Hamiltonian}) in the presence of the resonator field $\mathbf{E}^{}_{\mathrm{res}}(t)=\mathbf{E}^{}_{\mathrm{ac}}\cos(\omega t)$ and a parallel periodic bias field $\mathbf{E}^{}_{\mathrm{bias}}(t)$ with period $T^{}_{\mathrm{sw}}$ and amplitude $E^{}_{\mathrm{max}}$. In the specific experiment to be discussed below, this bias field is a symmetric triangular wave, as shown in Fig.~\ref{fig1}a). This bias field shifts the TLS bias energy, such that $\Delta(t)=\Delta(0)-2\mathbf{p}\cdot\mathbf{E}^{}_{\mathrm{bias}}(t)$. Under these assumptions, a number $n^{}_{\mathrm{TLS}}\propto P^{}_{0}\,pE^{}_{\mathrm{max}}$ of TLSs per unit volume are swept into resonance with the resonator field in each period of the bias field. In a single period, most of these TLSs experience two LZ transitions during which TLS dissipation is negligible for $\xi\gg\xi^{}_{1}=2\Gamma^{}_{1}/(\pi\Omega^{}_{\mathrm{R}0})$; the TLS dynamics in each resonance, occurring at time $t^{}_{0}$ for which the TLS energy splitting $\mathcal{E}(t)=\sqrt{\Delta^{2}(t)+\Delta^{2}_{0}}$ equals $\hbar\omega$ [Fig.~\ref{fig1}b)], is governed by the LZ Hamiltonian~\cite{Supp}
\begin{align}
\label{eq:LZ_Hamiltonian}&
\mathcal{H}^{}_{\mathrm{LZ}}(t)=\frac{1}{2}\left[v(t-t^{}_{0})\sigma^{}_{z}+\hbar\Omega^{}_{\mathrm{R}}\sigma^{}_{x}\right].
\end{align}  
Here, $\sigma^{}_{x}$ and $\sigma{}_{z}$ are the Pauli matrices in the diabatic basis $\{\ket{g,n},\ket{e,n-1}\}$ ($\ket{g}$ and $\ket{e}$ being the TLS ground and excited states, respectively, and $\ket{n}$ is a photon number state~\cite{Comment2}), and $v=v^{}_{0}\cos\eta\sqrt{1-(\Delta^{}_{0}/\hbar\omega)^{2}}$ is the TLS energy sweep rate, with $v^{}_{0}=2p\dot{E}^{}_{\mathrm{bias}}(t^{}_{0})$ the maximum sweep rate and $\eta$ the angle between the TLS dipole moment and the electric fields; the TLS Rabi frequency is $\Omega^{}_{\mathrm{R}}=\Omega^{}_{\mathrm{R}0}(\Delta^{}_{0}/\hbar\omega)\cos\eta$. Note that for the triangular bias field shown in Fig.~\ref{fig1}a), the maximum sweep rate is $|v^{}_{0}|=4pE^{}_{\mathrm{max}}/T^{}_{\mathrm{sw}}$.

To obtain the dielectric loss due to TLSs, we calculate the counting statistics of the number of photons absorbed by a single TLS. Within the full counting statistics formalism, the evolution operator describing a single coherent LZ transition is~\cite{SSN10}
\begin{align}
\label{eq:LZ_evolution}&
\mathcal{U}^{}_{\mathrm{LZ}}(k)=
\begin{pmatrix}
\sqrt{P} & e^{i\frac{k}{2}}e^{-i\psi}\sqrt{1-P} \\
-e^{-i\frac{k}{2}}e^{i\psi}\sqrt{1-P} & \sqrt{P}
\end{pmatrix},
\end{align} 
where $\psi$ is the Stokes phase, approaching 0 and $\pi/4$ in the adiabatic ($\xi\ll 1$) and non-adiabatic ($\xi\gg 1$) limits, respectively~\cite{SE91}. Note that a sign reversal of $v$ in the Hamiltonian~(\ref{eq:LZ_Hamiltonian}) corresponds to the transformation $\psi\rightarrow\pi-\psi$ in Eq.~(\ref{eq:LZ_evolution})~\cite{SSN10,Supp}. The counting field $k$ counts the number of photons absorbed by the TLS, with the factors $e^{-ik/2}$ and $e^{ik/2}$ corresponding to the absorption and emission of a photon. In Liouville space~\cite{Supp}, this evolution operator transforms into the superoperator $U^{}_{\mathrm{LZ}}(k)=\mathcal{U}^{}_{\mathrm{LZ}}(k)\otimes\left[\mathcal{U}^{}_{\mathrm{LZ}}(-k)\right]^{\ast}$.
	
In between two successive transitions, the TLS is out of resonance for a time interval $t$ and the dynamics of its density matrix $\rho$ is described by the Lindblad equation,
\begin{align}
\label{eq:Lindblad}&
\dot{\rho}=-\frac{i}{\hbar}\left[\mathcal{H}^{}_{\mathrm{TLS}},\rho\right]+\sum^{}_{i=\pm}\Gamma^{}_{i}\left(L^{}_{i}\rho L^{\dag}_{i}-\frac{1}{2}\{L^{\dag}_{i}L^{}_{i},\rho\}\right),
\end{align}  
where $\mathcal{H}^{}_{\mathrm{TLS}}(t)=(\mathcal{E}(t)/2)\sigma^{}_{z}$, $L^{}_{\pm}=\sigma^{}_{\pm}=(\sigma^{}_{x}\pm i\sigma^{}_{y})/2$ and $\Gamma^{}_{+}=\Gamma^{}_{\uparrow}$, $\Gamma^{}_{-}=\Gamma^{}_{\downarrow}$ are the transition rates between the TLS eigenstates. For simplicity, we assume no pure dephasing, such that the decoherence rate is $\Gamma^{}_{2}=\Gamma^{}_{1}/2$, where $\Gamma^{}_{1}=\Gamma^{}_{\uparrow}+\Gamma^{}_{\downarrow}$ is the relaxation rate. The corresponding evolution operator in Liouville space is~\cite{Supp}
\begin{align}
\label{eq:Evolution_operator_Lindblad}&
{\tiny U(t)=
\begin{pmatrix}
 \frac{\Gamma^{}_{\uparrow}}{\Gamma^{}_{1}}+\frac{\Gamma^{}_{\downarrow}}{\Gamma^{}_{1}}e^{-\Gamma^{}_{1}t} & 0 & 0 & \frac{\Gamma^{}_{\uparrow}}{\Gamma^{}_{1}}\left(1-e^{-\Gamma^{}_{1}t}\right) \\
 0 & e^{i\phi(t)-\Gamma^{}_{2}t} & 0 & 0 \\
 0 & 0 & e^{-i\phi(t)-\Gamma^{}_{2}t} & 0 \\
 \frac{\Gamma^{}_{\downarrow}}{\Gamma^{}_{1}}\left(1-e^{-\Gamma^{}_{1}t}\right) & 0 & 0 & \frac{\Gamma^{}_{\downarrow}}{\Gamma^{}_{1}}+\frac{\Gamma^{}_{\uparrow}}{\Gamma^{}_{1}}e^{-\Gamma^{}_{1}t}
\end{pmatrix}},
\end{align}
where $\phi(t)=\frac{1}{\hbar}\int^{t}_{0}\mathcal{E}(t')dt'$. The evolution of the density matrix after one period of the bias field is obtained as $\ket{\rho(T^{}_{\mathrm{sw}},k)}=U^{}_{\mathrm{sw}}(k)\ket{\rho(0)}$, where $\ket{\rho}=(\rho^{}_{00},\rho^{}_{01},\rho^{}_{10},\rho^{}_{11})^{T}$ is the ket representing the density matrix $\rho$ in Liuoville space~\cite{Supp}, and $U^{}_{\mathrm{sw}}(k)~=~U^{}_{\mathrm{LZ}}(\pi-\psi,k)U(t^{}_{2})U^{}_{\mathrm{LZ}}(\psi,k)U(t^{}_{1})$ with $T^{}_{\mathrm{sw}}=t^{}_{1}+t^{}_{2}$ (here we have used the fact that the sweep rate changes sign between consecutive transitions). The evolution after time $t=NT^{}_{\mathrm{sw}}$ is then
\begin{align}
\label{eq:Evolution}&
\ket{\rho(t,k)}=U^{N}_{\mathrm{sw}}(k)\ket{\rho(0)}.
\end{align}
	
The generating function for the statistics of the TLS photon absorption after time $t=NT^{}_{\mathrm{sw}}$ is given by
\begin{align}
\label{eq:Generating_function}
\chi(t,k)&=\mathrm{Tr}\left[\ket{\rho(t,k)}\right]=\mathrm{Tr}\left[U^{N}_{\mathrm{sw}}(k)\ket{\rho(0)}\right],
\end{align}
where the trace operation is defined as $\mathrm{Tr}\left[\ket{\rho}\right]\equiv\rho^{}_{00}+\rho^{}_{11}$. In particular, the number of photons absorbed by the TLS during time $t$ is given by the first moment $\braket{N^{}_{\mathrm{ph}}(t)}=-i\frac{\partial\chi(t,k)}{\partial k}\big|^{}_{k=0}$. For $k=0$ there should be a stationary solution to Eq.~(\ref{eq:Evolution}), meaning that one of the eigenvalues $\lambda^{}_{1}$ of $U^{}_{\mathrm{sw}}(k=0)$ equals unity, whereas $|\lambda^{}_{j}|<1$ for $j=2,3,4$. As a result, in the limit $t\rightarrow\infty$ only the mode with eigenvalue $\lambda^{}_{1}=1$ will contribute, and after some algebra we obtain the photon absorption rate per TLS~\cite{Supp},
\begin{align}
\label{eq:photon_absorption_rate_TLS}&
\gamma^{}_{\mathrm{abs}}=\lim^{}_{t\rightarrow\infty}\frac{\braket{N^{}_{\mathrm{ph}}(t)}}{t}=-\frac{i}{T^{}_{\mathrm{sw}}}\bra{g^{}_{1}}\frac{dU^{}_{\mathrm{sw}}}{dk}\bigg|^{}_{k=0}\ket{v^{}_{1}},
\end{align} 
where $\bra{g^{}_{1}}$ and $\ket{v^{}_{1}}$ are the left and right eigenvectors of $U^{}_{\mathrm{sw}}(k=0)$ corresponding to the eigenvalue $\lambda^{}_{1}=1$. The total photon absorption rate per unit volume is $\Gamma^{}_{\mathrm{abs}}=n^{}_{\mathrm{TLS}}\gamma^{}_{\mathrm{abs}}\propto P^{}_{0}\,pE^{}_{\mathrm{max}}\gamma^{}_{\mathrm{abs}}$. Comparing the power dissipation density $P^{}_{\mathrm{dis}}=-\hbar\omega\Gamma^{}_{\mathrm{abs}}$ with $P^{}_{\mathrm{dis}}=-\frac{1}{2}\omega\epsilon''E^{2}_{\mathrm{ac}}$, we obtain the expression for the loss tangent
\begin{align}
\label{eq:loss_tangent}&
\tan\delta=\frac{\epsilon''}{\epsilon'}=\frac{2\hbar\Gamma^{}_{\mathrm{abs}}}{\epsilon E^{2}_{\mathrm{ac}}}=\frac{2p^{2}\Gamma^{}_{\mathrm{abs}}}{\epsilon\hbar\Omega^{2}_{\mathrm{R}0}},
\end{align}
where $\epsilon'$ and $\epsilon''$ are the real and imaginary parts of the dielectric constant.	
	
The general expression for $\gamma^{}_{\mathrm{abs}}$ is somewhat complicated, see Eq.~(26) in the supplementary material~\cite{Supp}. We now consider the experimentally relevant regime $k^{}_{\mathrm{B}}T\ll\hbar\omega$, for which $\Gamma^{}_{1}\approx\Gamma^{}_{\downarrow}$ ($\Gamma^{}_{\uparrow}\approx 0$), and analyze the expression for $\gamma^{}_{\mathrm{abs}}$ in simple limits. We first consider the incoherent limit $\Gamma^{}_{1}T^{}_{\mathrm{sw}}\gg 1$, which in terms of the dimensionless sweep rate $\xi$ can be expressed as $\xi\ll\xi^{}_{2}$, with $\xi^{}_{2}\equiv 8pE^{}_{\mathrm{max}}\Gamma^{}_{1}/(\pi\hbar\Omega^{2}_{\mathrm{R}0})$. In this limit we obtain $\gamma^{}_{\mathrm{abs}}\approx 2\left(1-P\right)/T^{}_{\mathrm{sw}}$. Equation~(\ref{eq:loss_tangent}) then gives the universal behavior discussed in Refs.~\cite{BAL13,KMS14}, namely $\tan\delta/\tan\delta^{}_{0}\approx 1$ in the non-adiabatic limit $\xi\gg 1$, and $\tan\delta/\tan\delta^{}_{0}\approx\xi$ for $\xi^{}_{1}\ll\xi\ll 1$~\cite{Supp}. Thus, the results of Refs.~\cite{BAL13,KMS14} are reproduced if subsequent LZ transitions are incoherent such that TLSs start from the ground state at each transition. We note that the regime $\xi<\xi^{}_{1}=2\Gamma^{}_{1}/(\pi\Omega^{}_{\mathrm{R}0})$, in which dissipation occurs within a single LZ transition, has to be treated separately. In this limit the loss approaches the saturation limit $\tan\delta/\tan\delta^{}_{0}=1/\sqrt{1+(\Omega^{}_{\mathrm{R}0}/\Gamma^{}_{1})^{2}}$, as studied numerically in Ref.~\cite{BAL13}. As mentioned above, in this work we concentrate on the regime $\xi>\xi^{}_{1}$, where dissipation within a single transition can be safely neglected, and consider the effect of dissipation between transitions.
	
In the coherent regime $\Gamma^{}_{1}T^{}_{\mathrm{sw}}\ll 1$ or $\xi\gg\xi^{}_{2}$, TLSs experience $M=\left(\Gamma^{}_{1}T^{}_{\mathrm{sw}}\right)^{-1}=\xi/\xi^{}_{2}\gg 1$ multiple coherent transitions. In the non-adiabatic regime $\xi\gg 1$, where the probability $1-P$ for photon absorption\textbackslash emission in a single transition is small, the interference between multiple transitions is constructive for $\phi^{}_{1}+\phi^{}_{2}=2\pi n$~\cite{Supp}, where $n$ is an integer and $\phi^{}_{1,2}$ are the dynamical phases accumulated between successive transitions. This gives rise to a resonance in $\gamma^{}_{\mathrm{abs}}$ as a function of the phases, whose width in the non-adiabatic regime $\xi\gg 1$ is $\delta\phi\propto M^{-1}$ for $M^{2}\left(1-P\right)<1$ and $\delta\phi\propto\sqrt{1-P}$ for $M^{2}\left(1-P\right)>1$~\cite{Supp}. The contribution to $\gamma^{}_{\mathrm{abs}}$ of TLSs out of resonance (corresponding to destructive interference~\cite{Supp}) is $\gamma^{\mathrm{non-res}}_{\mathrm{abs}}\propto\Gamma^{}_{1}\left(1-P\right)=\Gamma^{}_{1}/\xi$, with weak dependence on $\phi^{}_{1}$ and $\phi^{}_{2}$. Below we concentrate on the contribution of the resonance, which dominates over that of the off-resonance part.

To obtain the loss tangent due to an ensemble of TLSs, one has to compute the total absorption rate per unit volume [see Eq.~(\ref{eq:loss_tangent})], $\Gamma^{}_{\mathrm{abs}}$, by averaging $\gamma^{}_{\mathrm{abs}}$ over the distribution of TLSs and the orientation of their dipole moments, as described in the supplementary material~\cite{Supp}. This is a complicated procedure~\cite{Supp}, and instead we choose to concentrate on the main effect of the ensemble of TLSs relevant to the interference discussed above, which is the distribution of the phases $\phi^{}_{1}$ and $\phi^{}_{2}$. It is plausible to assume that the wide, random distribution of TLS parameters translates into an approximately homogeneous distribution of $\phi^{}_{1}$ and $\phi^{}_{2}$. We thus neglect the distribution of $\Delta^{}_{0}$, $p$ and $\eta$ in all other quantities, such as the sweep rate, the Rabi frequency, the relaxation rate and the stokes phase, and set $t^{}_{1}=t^{}_{2}=T^{}_{\mathrm{sw}}/2$ (the qualitative results are not sensitive to the latter choice). The absorption rate per TLS, $\gamma^{}_{\mathrm{abs}}$, is then a function of $\xi$, $\xi^{}_{2}$, $\phi^{}_{1}$ and $\phi^{}_{2}$~\cite{Supp}. Two different behaviors of the loss tangent in the coherent regime are expected for $\xi^{}_{2}<1$ and $\xi^{}_{2}>1$. 

\begin{figure}[ht!]
\includegraphics[width=0.47\textwidth,height=5.3cm]{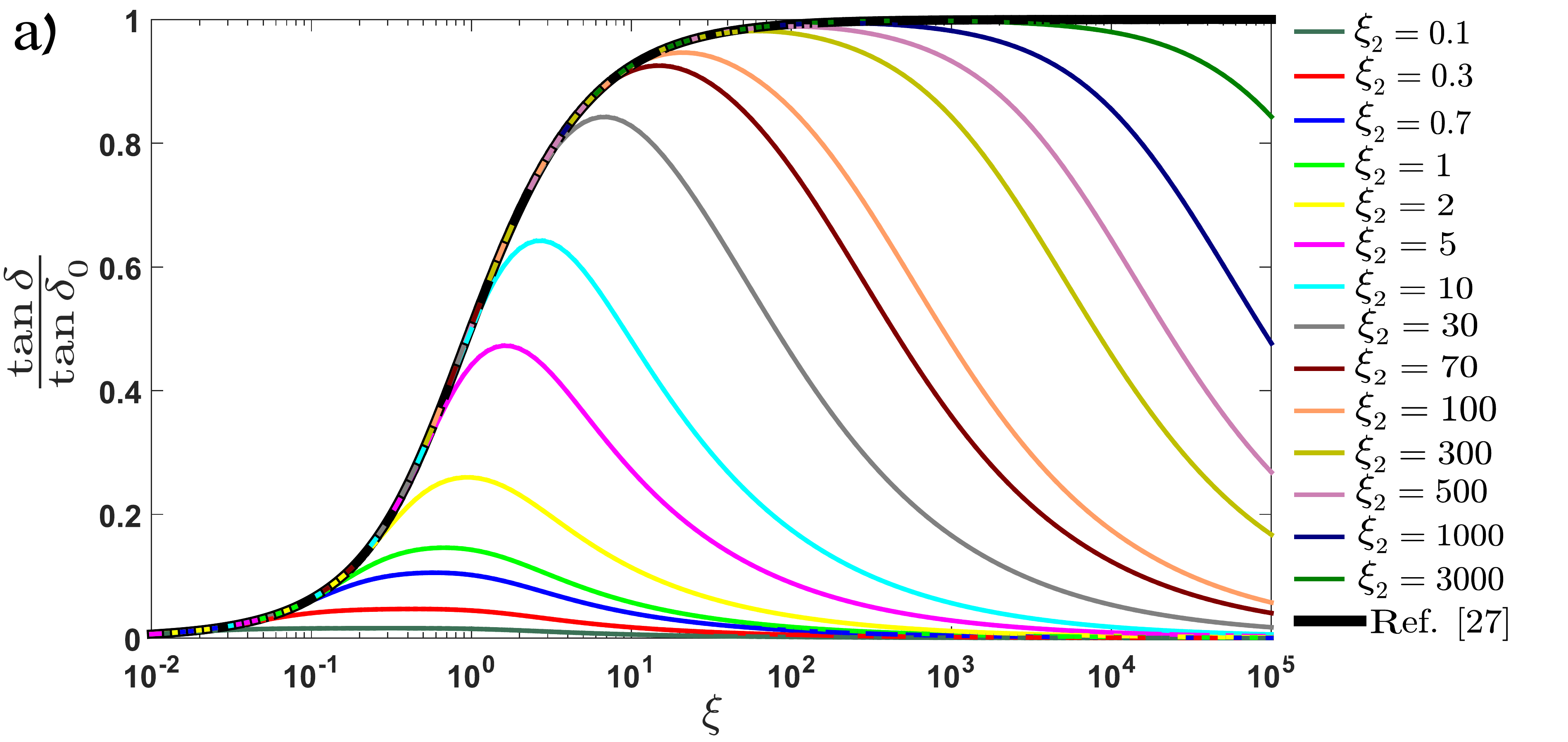}
\includegraphics[width=0.47\textwidth,height=5.3cm]{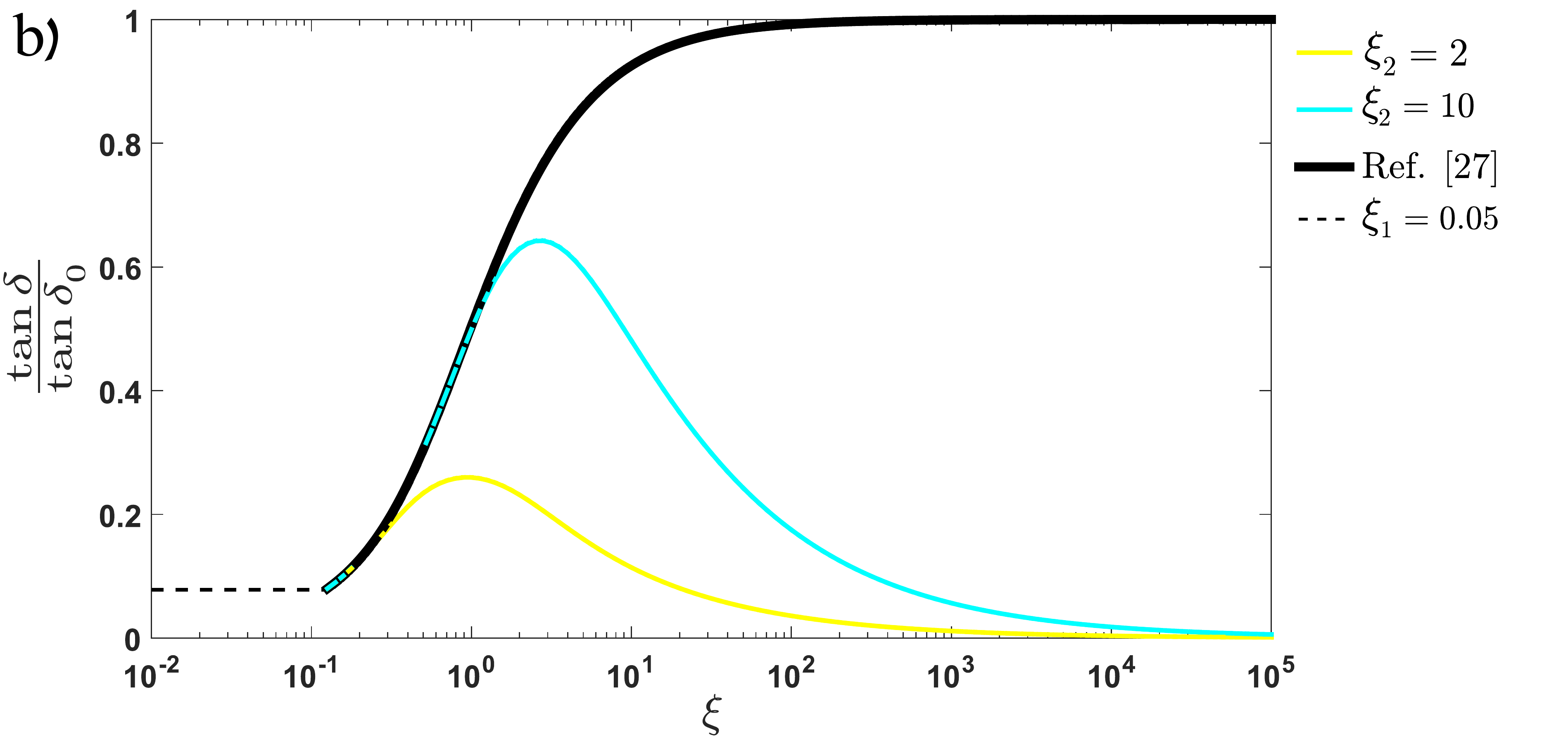}
\includegraphics[width=0.47\textwidth,height=5.3cm]{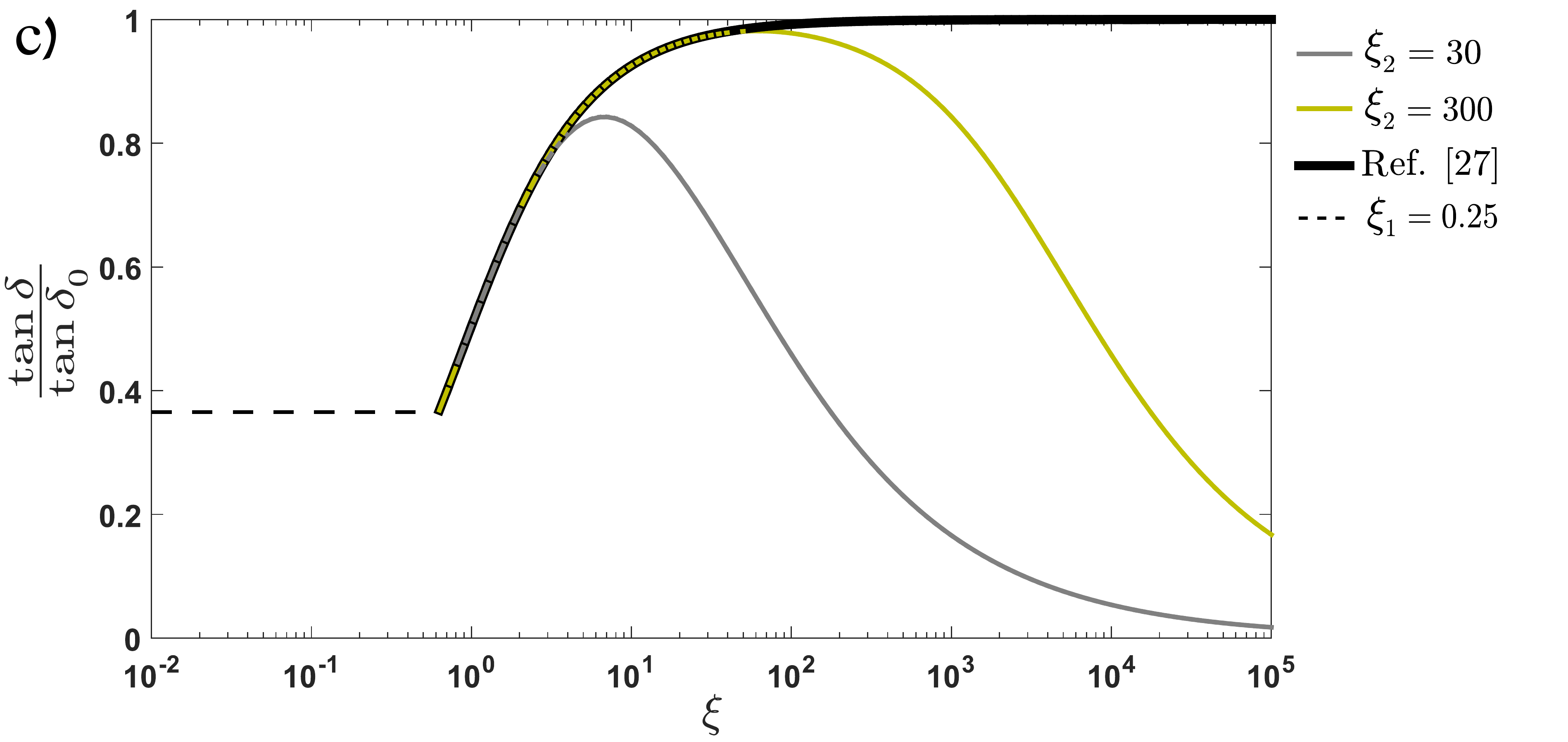}
\caption{Theoretical results for the loss tangent due to TLSs, normalized by the intrinsic low-power loss tangent $\tan\delta^{}_{0}=\pi P^{}_{0}p^{2}/(3\epsilon)$, as a function of the dimensionless sweep rate $\xi\equiv 2|v^{}_{0}|/(\pi\hbar\Omega^{2}_{\mathrm{R}0})$ for various values of $\xi^{}_{2}\equiv 8pE^{}_{\mathrm{max}}\Gamma^{}_{1}/(\pi\hbar\Omega^{2}_{\mathrm{R}0})$, as indicated in the legend. The results are obtained by a numerical average of the photon absorption rate per TLS, $\braket{\gamma^{}_{\mathrm{abs}}}$, over a homogeneous distribution of the phases $\phi^{}_{1}$ and $\phi^{}_{2}$. a) Calculation in the limit $\xi^{}_{1}=2\Gamma^{}_{1}/(\pi\Omega^{}_{\mathrm{R}0})\rightarrow0$ and $pE^{}_{\mathrm{max}}/(\hbar\Gamma^{}_{1})\rightarrow\infty$ such that $\xi^{}_{2}=8pE^{}_{\mathrm{max}}\Gamma^{}_{1}/(\pi\hbar\Omega^{2}_{\mathrm{R}0})=2\pi\xi^{2}_{1}pE^{}_{\mathrm{max}}/(\hbar\Gamma^{}_{1})$ is finite, corresponding to full saturation at zero sweep rate. b) Modification of the results for the case of partial saturation at zero sweep rate (finite $\xi^{}_{1}$), with $\xi^{}_{1}=0.05$ and c) $\xi^{}_{1}=0.25$. The loss in a) is cut by the saturation value $\tan\delta/\tan\delta^{}_{0}=1/\sqrt{1+(\Omega^{}_{\mathrm{R}0}/\Gamma^{}_{1})^{2}}$ (horizontal dashed line), to which it approaches for $\xi\lesssim\xi^{}_{1}$~\cite{BAL13,KMS14}. Due to the decoupling effect at high sweep rates the loss reduces below its value in the absence of a periodic bias field ($\xi=0$).}
\label{fig2}
\end{figure}
	
For $\xi^{}_{2}<1$, the regime $\xi^{}_{2}<\xi\ll 1$ is coherent ($\xi>\xi^{}_{2}$) and adiabatic ($\xi\ll 1$), meaning that photons are absorbed and re-emitted by the TLSs with high probability. The photons are thus dissipated at the relaxation rate of the TLSs, so that $\gamma^{}_{\mathrm{abs}}\propto\Gamma^{}_{1}$~\cite{Supp}. This gives rise to constant loss tangent, $\tan\delta\propto\xi^{}_{2}$. In the non-adiabatic regime $\xi^{}_{2}<1\ll\xi$ the resonance width is $\delta\phi\propto\sqrt{1-P}$ (since $M^{2}\left(1-P\right)\approx\xi/\xi^{2}_{2}\gg 1$ in this regime). If $\phi^{}_{1}$ and $\phi^{}_{2}$ are nearly homogeneously distributed, the contribution of this resonance to the photon absorption rate is $\gamma^{\mathrm{res}}_{\mathrm{abs}}\propto\gamma^{}_{\mathrm{abs}}(\phi^{}_{1}=-\phi^{}_{2})\cdot\delta\phi\propto\Gamma^{}_{1}\sqrt{1-P}$. Hence, the loss tangent decreases as $\xi^{-1/2}$.
	
For $\xi^{}_{2}>1$, the loss tangent follows the universal curve of Ref.~\cite{BAL13} up to $\xi\sim\xi^{}_{2}$. For $\xi\gg\xi^{2}_{2}$ the resonance width is again $\delta\phi\propto\sqrt{1-P}$, and the corresponding contribution of this resonance to the loss tangent is again $\propto\xi^{-1/2}$. In the crossover region $\xi^{}_{2}<\xi\ll\xi^{2}_{2}$ the resonance width is $\delta\phi\propto M^{-1}$, giving rise to the photon absorption rate $\gamma^{\mathrm{res}}_{\mathrm{abs}}\approx\gamma^{}_{\mathrm{abs}}(\phi^{}_{1}=-\phi^{}_{2})\cdot\delta\phi\propto\Gamma^{}_{1}M\left(1-P\right)$, which depends weakly on $\xi$. Table~\ref{tab:regimes} summarizes the qualitative behavior of the loss tangent in various regimes.
\begin{table*}[]
	\begin{tabular}{|c|c|c|c|c|c|c|c|}
		\hline
		\multicolumn{1}{|c|}{} & \multicolumn{3}{|c|}{$\xi^{}_{2}<1$} & \multicolumn{4}{c|}{$\xi^{}_{2}>1$} \\ \hline
	& $\xi<\xi^{}_{2}$ & $\xi^{}_{2}<\xi\ll 1$ & $\xi\gg 1$ & $\xi\ll 1$ & $1\ll\xi<\xi^{}_{2}$ & $\xi^{}_{2}<\xi\ll\xi^{2}_{2}$ & $\xi\gg\xi^{2}_{2}$ \\ \hline
	$\tan\delta/\tan\delta^{}_{0}$ & $\propto\xi$ & $\propto\xi^{}_{2}$ & $\propto\xi^{}_{2}/\sqrt{\xi}$ & $\propto\xi$ & $\propto1$ & $\propto1$ & $\propto\xi^{}_{2}/\sqrt{\xi}$ \\ \hline
	\end{tabular}
\caption{ Qualitative behavior of the normalized loss tangent $\tan\delta/\tan\delta^{}_{0}$ in various regimes.}
\label{tab:regimes}
\end{table*}

In Fig.~\ref{fig2}a) we show the results for the loss tangent obtained by a numerical average of the absorption rate over the homogeneous distribution of $\phi^{}_{1}$ and $\phi^{}_{2}$. One readily observes the qualitative limits discussed above. The results in Fig.~\ref{fig2}a) are obtained for the limit $\xi^{}_{1}=2\Gamma^{}_{1}/(\pi\Omega^{}_{\mathrm{R}0})\rightarrow 0$, such that TLSs are fully saturated at zero sweep rate (i.e., we take the limits $\xi^{}_{1}\rightarrow0$ and $pE^{}_{\mathrm{max}}/(\hbar\Gamma^{}_{1})\rightarrow\infty$ such that $\xi^{}_{2}=8pE^{}_{\mathrm{max}}\Gamma^{}_{1}/(\pi\hbar\Omega^{2}_{\mathrm{R}0})=2\pi\xi^{2}_{1}pE^{}_{\mathrm{max}}/(\hbar\Gamma^{}_{1})$ is finite). This shows how the universal curve discussed in Ref.~\cite{BAL13} (solid black curve in Fig.~\ref{fig2}a)) is modified due to multiple coherent transitions. Note that under this assumption the loss at high sweep rates cannot reduce below the vanishing loss at $\xi=0$.
	
In order to relate directly to experiment, we note that for finite $\xi^{}_{1}$ the loss approaches the saturation limit $\tan\delta/\tan\delta^{}_{0}=1/\sqrt{1+(\Omega^{}_{\mathrm{R}0}/\Gamma^{}_{1})^{2}}=1/\sqrt{1+(2/\pi\xi^{}_{1})^2}$ for $\xi<\xi^{}_{1}$~\cite{BAL13,KMS14}. In Figs.~\ref{fig2}b) and ~\ref{fig2}c) we show the theoretical results expected for finite values of $\xi^{}_{1}$. For each value of $\xi^{}_{2}>\xi^{}_{1}$ (which translates to a given value of $pE^{}_{\mathrm{max}}/(\hbar\Gamma^{}_{1})=\xi^{}_{2}/(2\pi\xi^{2}_{1})$), the results in Fig.~\ref{fig2}a) describe the loss at $\xi>\xi^{}_{1}$. For $\xi<\xi^{}_{1}$ we cut these results by the horizontal lines corresponding to the value of the loss at the stationary saturation limit $\xi=0$. In practice, for $\xi\lesssim\xi^{}_{1}$ the loss tangent reduces to its $\xi=0$ value monotonically with decreasing $\xi$, as was studied numerically in Refs.~\cite{BAL13,KMS14}. Thus, in the regime $\xi^{}_{1}<\xi<\xi^{}_{2}$ the loss tangent is described by the universal curve discussed in Ref.~\cite{BAL13}, whereas it becomes non-universal for $\xi<\xi^{}_{1}$ (due to dissipation within a single transition) or $\xi>\xi^{}_{2}$ (due to coherent multiple transitions). As seen in Figs.~\ref{fig2}b) and ~\ref{fig2}c), for finite $\xi^{}_{1}$ one expects the loss at high sweep rates ($\xi\gg\xi^{}_{2}$) to decrease below its value at $\xi=0$. All the main qualitative features of our theoretical results are observed experimentally. This includes also the saturation of the loss at $\xi<\xi^{}_{1}$, and to some extent the decrease below this value at large sweep rates (see Fig.~\ref{fig4} below).

We stress that the decrease of the loss at the coherent and non-adiabatic regime $\xi\gg\max\{1,\xi^{}_{2}\}$ is a result of {\it interference between $M$ coherent LZ transitions}, which reduce the photon absorption probability. To see this, consider $N$ identical TLSs of which $N^{}_{\mathrm{g}}(t)$ and $N^{}_{\mathrm{e}}(t)$ occupying the ground and excited states, respectively. In a classical approach~\cite{Comment1}, one can write a rate equation for $N^{}_{\mathrm{e}}(t)$,
\begin{align}
\label{eq:rate_equation}
\dot{N}^{}_{\mathrm{e}}&=\gamma(N^{}_{\mathrm{g}}-N^{}_{\mathrm{e}})-\Gamma^{}_{\downarrow}N^{}_{\mathrm{e}}+\Gamma^{}_{\uparrow}N^{}_{\mathrm{g}}\nonumber\\
&=\gamma(N-2N^{}_{\mathrm{e}})-\Gamma^{}_{1}N^{}_{\mathrm{e}}+\Gamma^{}_{\uparrow}N,
\end{align}
where $\gamma=2\left(1-P\right)/T^{}_{\mathrm{sw}}$ is the photon emission and absorption rate in a single LZ transition. The steady state solution is $N^{}_{\mathrm{e}}=N(\gamma+\Gamma^{}_{\uparrow})/\left(2\gamma+\Gamma^{}_{1}\right)$ and the corresponding photon absorption rate per TLS is
\begin{align}
\label{eq:photon_absorption_rate_classical}
\gamma^{}_{\mathrm{abs}}=\frac{\gamma\left(N-2N^{}_{\mathrm{e}}\right)}{N}=\frac{\Gamma^{}_{\downarrow}-\Gamma^{}_{\uparrow}}{\Gamma^{}_{1}}\frac{\gamma}{1+2\gamma/\Gamma^{}_{1}}.
\end{align}
Since $(\Gamma^{}_{\downarrow}-\Gamma^{}_{\uparrow})/\Gamma^{}_{1}=\tanh(\hbar\omega/2k^{}_{\mathrm{B}}T)$ equals unity at low temperatures, we obtain $\gamma^{}_{\mathrm{abs}}\approx\gamma$ for $\gamma\ll\Gamma^{}_{1}$ (or $M(1-P)\ll 1$) and $\gamma^{}_{\mathrm{abs}}\approx\Gamma^{}_{1}/2$ for $\gamma\gg\Gamma^{}_{1}$ (or $M(1-P)\gg 1$). The first limit corresponds to the result of Refs.~\cite{KMS14,BAL13} and the second limit corresponds to a constant loss tangent $\tan\delta\propto\xi^{}_{2}$, as we find above in the regime $\xi^{}_{2}<\xi\ll 1$. Therefore, a classical approach based on independent transitions does not capture the physics of the fast sweep regime, which exhibits a decreasing loss with increasing sweep rate for $\xi>\xi^{}_{2}$.

\subsection{Experiment}  \label{Experiment}
In our experiment, we study TLS in deposited aluminum oxide by using it as the dielectric in lumped-element LC-resonators. This material is highly relevant for superconducting quantum processors, because it is used for tunnel barriers in Josephson junctions of qubits and also forms naturally on circuit wiring after air exposure. However, any depositable dielectric can in principle be studied with this method.

\begin{figure}[htb!]
\includegraphics[width=0.99\columnwidth,height=5cm]{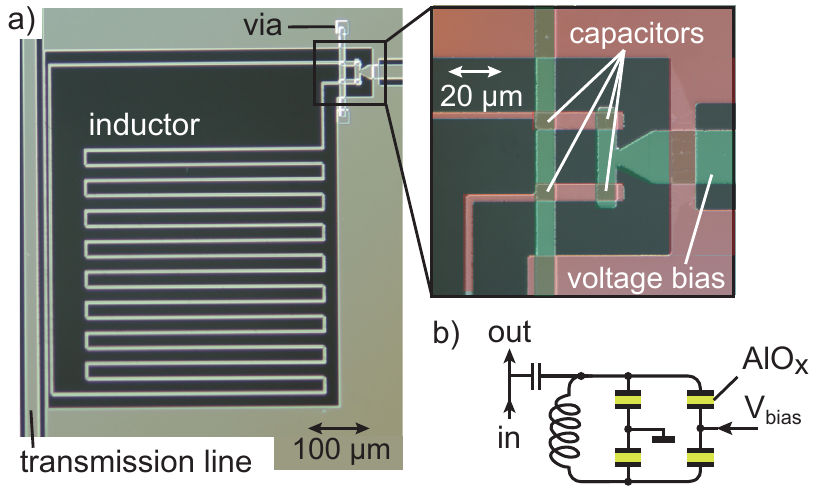} 
\caption{a) Photograph of a lumped-element resonator consisting of a capacitively terminated meandering inductor. The (colorized) inset shows a zoom of the four capacitors between bottom (red) and top layers (green), which are separated by 25nm-thick amorphous AlO$_x$. b) Circuit schematic. The electric field in the capacitor dielectric is controlled by an applied bias voltage $V^{}_{\mathrm{bias}}$.}
\label{fig3}
\end{figure}

Figure~\ref{fig3} shows a sample resonator structured by optical lithography from superconducting aluminum on a sapphire substrate. Following experiments by Khalil et al.~\cite{KMS14}, the capacitances are designed as bridges consisting of four equal Al/AlOx/Al capacitors. Hereby, an electric bias field can be applied to the dielectric. In addition, our setup allows for mechanical TLS tuning by controlling the strain in the sample material with a piezo actuator~\cite{LJ15}. Each chip contains 8 slightly different resonators that are coupled to a common transmission line, and is installed in a well-shielded and heavily filtered cryogenic setup that allows for measurements in the single-photon regime at sample temperatures of 30 mK~\cite{BJD17}. All capacitors contain a 25-nm thick layer of amorphous AlO$_x$ that is deposited in a Plassys system by eBeam-evaporation of aluminum in a low-pressure oxygen atmosphere. Further details on the setup and fabrication are found in~\cite{Supp}.

We characterize the total dielectric loss tangent $\tan\delta\equiv 1/Q^{}_{i}$ by recording resonance curves using a network analyzer and extracting the internal quality factor $Q^{}_{i}$ using a standard fit procedure~\cite{PS15}. In particular, we study this loss while a triangular voltage signal $V^{}_{\mathrm{bias}}(t)$ is applied as a bias to the sample dielectric. This results in a sweep rate $v^{}_{0}=p\dot{V}^{}_{\mathrm{bias}}/d$, where $d=25\,$nm is the distance between the capacitor plates, considering that due to the design only half the voltage drops at each capacitor. The shortest periods in our experiment are $10\,$ns, such that $T^{}_{\mathrm{sw}}\gg 2\pi/\omega$, where $\omega\approx 2\pi\times 7\,$GHz is the resonance frequency of the resonator. Resonant transitions due to the bias field can therefore be safely neglected. The highest bias field amplitude is $E^{}_{\mathrm{max}}=0.9\,$MV/m, which allows us to apply a bias field rate $\dot{E}^{}_{\mathrm{bias}}=2E^{}_{\mathrm{max}}/T^{}_{\mathrm{sw}}$ up to $1.8\cdot 10^{14}\,\mathrm{V}/(\mathrm{m}\cdot\mathrm{s})$. For typical values of the dipole moment of TLSs in AlO$_x$, $p\approx 0.5\,e\AA$, this corresponds to a maximum sweep rate of $|v^{}_{0}|/h\approx 2\cdot 10^{9}\,$GHz/s. The adiabatic condition $v^{}_{0}/(\hbar\omega^{2})\ll 1$ thus holds, justifying the assumption that the bias field changes the energy splitting of the TLS adiabatically. We also note that as in Refs.~\cite{BJD17,SB15,SB16}, the dielectric volume of the resonator has been chosen such that on average there is roughly one TLS in resonance with the resonator in the absence of the bias field. By applying the bias field all TLSs within the energy window $pE^{}_{\mathrm{max}}$ around $\hbar\omega$ are swept into resonance and contribute to the loss. In our experiment $pE^{}_{\mathrm{max}}$ is in the range $1-10\,$GHz, hence $\sim100-1000$ TLSs contribute to the loss and averaging is proper.

\begin{figure}[htb!]
	\includegraphics[width=0.99\columnwidth,height=12cm]{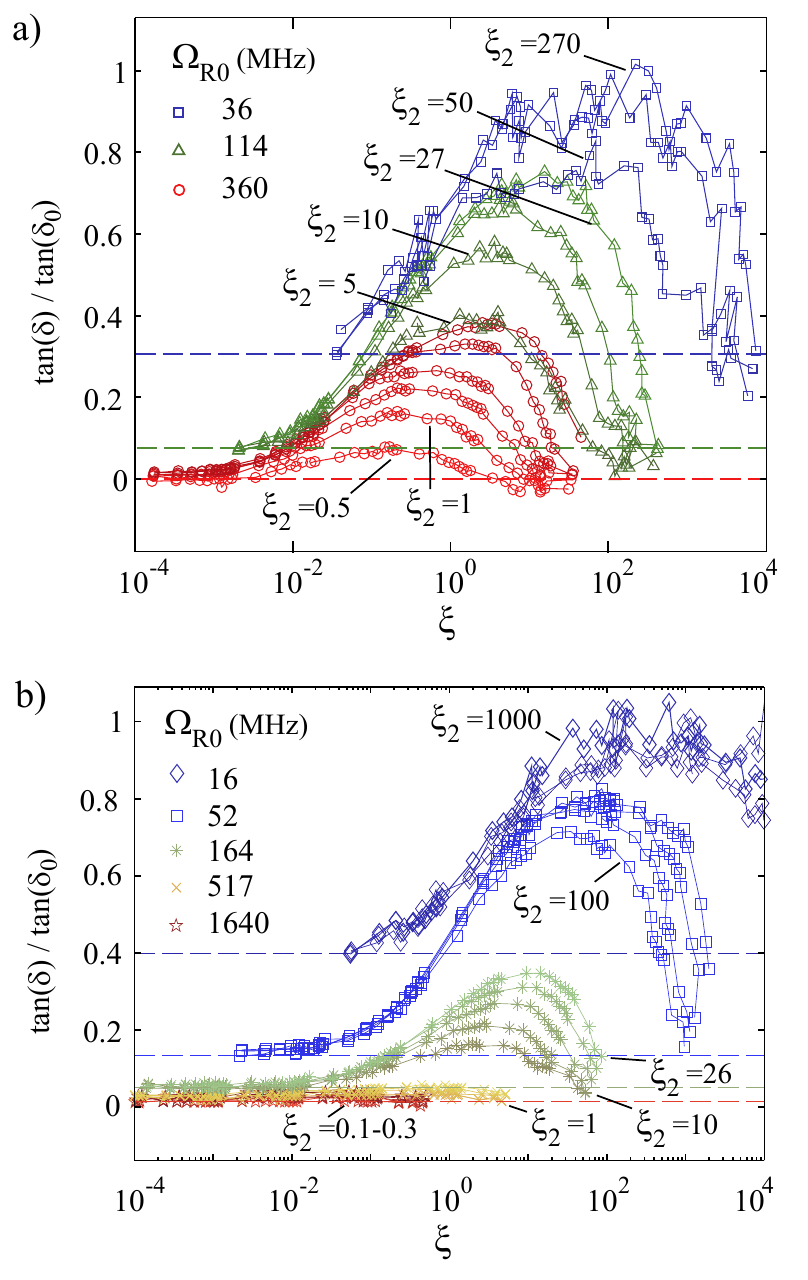} 
	\caption{Normalized dielectric loss as a function of the dimensionless sweep rate $\xi$ for various values of $\xi^{}_{2}$ which were set by the applied microwave power (coloured curve families) and the amplitude of the bias field (varied in the range $89.5-806$\,kV/m) a) for resonator 1 and b) resonator 2. For each family of curves obtained with the same input power, the legend shows the corresponding value of $\Omega^{}_{\mathrm{R}0}$ calculated at $\xi=0$. Horizontal dashed lines indicate the resonator loss when it is driven at the power levels of the curve families without applied bias field. Negative loss values are within the standard deviation.} 
	\label{fig4}
\end{figure}

Figure~\ref{fig4} shows the measured dielectric loss tangent in two different resonators as a function of the dimensionless sweep rate $\xi=2|v^{}_{0}|/(\pi\hbar\Omega^{2}_{\mathrm{R}0})$. Each curve is obtained by varying the period $T^{}_{\mathrm{sw}}$ of the bias field, keeping its amplitude $E^{}_{\mathrm{max}}$ and the input power $P^{}_\mathrm{in}$ fixed. To calculate $\xi$, the maximum Rabi frequency is computed as $\Omega^{}_{\mathrm{R}0}=pE^{}_\mathrm{ac}/\hbar=\sqrt{\left(p^{2}P^{}_\mathrm{in}Q^{2}_{l}\right)/\left(\hbar^{2}\omega CQ^{}_{\mathrm{c}}d^{2}\right)}$, where $Q^{}_{l}$, $Q^{}_{c}$ are the measured loaded and coupling quality factors, $C$ is the total resonator capacitance, and a typical value of $p=0.5\,$e$\AA$ is used~\cite{MC18}. Note that for a given $P^{}_{\mathrm{in}}$, the value of $E^{}_{\mathrm{ac}}$ (and thus $\Omega^{}_{\mathrm{R0}}$) depends on the resonator loss (and thus on $\xi$). The values of $\xi$ in the horizontal axes of Fig.~~\ref{fig4} take this dependence into account.

To compare the experimental results with our theory, Fig.~\ref{fig4} shows the loss due to TLSs, obtained by subtracting the background loss at the saturation regime (large powers) at $\xi=0$. We further normalize the resulting loss tangent by the intrinsic loss tangent $\tan\delta^{}_{0}$ (see~\cite{Supp} for saturation curves of the resonators and for values of the background and intrinsic TLS loss tangent). In addition, we estimate the values of the parameter $\xi^{}_{2}=8pE^{}_{\mathrm{max}}\Gamma^{}_{1}/(\pi\hbar\Omega^{2}_{\mathrm{R}0})$ for selected curves. For this purpose, we use the value of $\Omega^{}_{\mathrm{R0}}$ at $\xi=0$ and set $p=0.5\,e\AA$ and $\Gamma^{}_{1}=1\,$MHz, in accordance with TLS dipole moments and relaxation rates observed in AlO$_x$~\cite{MC18,SY10,LJ10,LJ16}. Note that this is an approximation, since the value of $\xi^{}_{2}$ is not constant for measurement at a fixed power (due to the loss dependence of the Rabi frequency discussed above). For both resonators, the qualitative agreement with the theoretical prediction of Figs.~\ref{fig2} is excellent. By tuning the input power, and therefore varying $\Omega^{}_{\mathrm{R0}}$, one can change the parameter $\xi^{}_{2}$ by several orders of magnitude to obtain the different behaviors shown in Figs.~\ref{fig2}. Variation of $E^{}_{\mathrm{max}}$ then weakly tunes the value of $\xi^{}_{2}$ in each regime. For $\xi^{}_{2}\lesssim 1$ one observes wide peaks which become more pronounced for $1\lesssim\xi^{}_{2}<100$. For $\xi^{}_{2}>100$ these peaks become the universal plateau as in Fig.~\ref{fig4}b), followed by the reduction in loss. Note that for $\xi^{}_{2}>1$ the loss starts decreasing at $\xi\approx\xi^{}_{2}$, in agreement with the theoretical prediction of Fig.~\ref{fig2}a). Unfortunately, comparison of the functional form of this decrease with the power low $\tan\delta\propto\xi^{-1/2}$ predicted by our theory is impossible, both because there is almost no data at the regime $\xi>\xi^{2}_{2}$ and because of the dependence of $\Omega^{}_{\mathrm{R}0}$ on $\xi$, not taken into account by the theory. We also notice that resonator 1 [Fig.~\ref{fig4}a)] provides some evidence that the loss at high sweep rates can reduce below its value at $\xi=0$ (no bias field). This is seen for the green and blue curve families for which the TLSs are not fully saturated at $\xi=0$.

\section{The single-photon regime and correspondence to a qubit coupled to resonant TLSs} \label{Qubits}
To examine whether the effect discussed above is also applicable in the single-photon regime, we consider now a quantized single-mode cavity field $\mathbf{E}^{}_{\mathrm{res}}(t)=\hat{\mathbf{e}}\sqrt{\hbar\omega/\epsilon^{}_{0}V}(ae^{-i\omega t}+a^{\dag}e^{i\omega t})\sin(kz)$ ($\hat{\mathbf{e}}$ is a polarization unit vector, $\epsilon^{}_{0}$ the vacuum permittivity, $V$ the resonator volume, $k$ the wave vector, and $a^{\dag}$, $a$ the photon creation and annihilation operators) propagating along the $z$-axis and interacting with a set of near-resonant TLSs. After neglecting the longitudinal coupling and applying the rotating wave approximation, the corresponding Hamiltonian is
\begin{align}
\label{eq:TLS_field_Hamiltonian}
\mathcal{H}=&\frac{1}{2}\sum^{}_{i}\mathcal{E}^{}_{i}\sigma^{i}_{z}+\hbar\omega a^{\dag}a+\sum^{}_{i}g^{}_{i}\left(\sigma^{i}_{+}a+\sigma^{i}_{-}a^{\dag}\right),
\end{align}
where $g^{}_{i}=-p(\Delta^{}_{0}/\mathcal{E}^{}_{i})\sqrt{\hbar\omega/\epsilon^{}_{0}V}\sin(kz^{}_{i})$ and $\sigma^{}_{\pm}=(\sigma^{}_{x}\pm i\sigma^{}_{y})/2$. As the last term couples different TLSs via the quantized cavity field, the assumption of independent TLSs cannot be invoked as in the case of a classical field discussed above (which corresponds to the substitution $2g^{}_{i}\sqrt{n^{}_{\mathrm{ph}}}=\Omega^{}_{\mathrm{R},i}$). For $\braket{n^{}_{\mathrm{ph}}}\gg 1$, each TLS feels the same classical field in every transition, but for $\braket{n^{}_{\mathrm{ph}}}\sim 1$ the dynamics of each transition depends on previous transitions of other TLSs. In this regime, calculation of the probability for an absorption of a single photon involves a consideration of multiple emissions and absorptions by an ensemble of TLSs, and thus the interference between many more paths than in the above analysis, where the coherent evolution of a single TLS was considered. It is expected, however, that just as in the case of independent TLSs discussed above, the random distribution of TLSs leads to random distribution of phases accumulated between consecutive transitions. As a result, there will be no preference for some resonant paths that involve emissions and absorptions of multiple TLSs, and most paths will interfere destructively, thus justifying an independent treatment of each TLS. In this case, each TLS is described by a Jaynes-Cummings Hamiltonian~\cite{BM11},  
\begin{align}
\label{eq:Jaynes_Cummings_Hamiltonian}&
\mathcal{H}=\frac{1}{2}\mathcal{E}\sigma^{}_{z}+\hbar\omega a^{\dag}a+g\left(\sigma^{}_{+}a+\sigma^{}_{-}a^{\dag}\right),
\end{align}
which reduces to the LZ Hamiltonian in the vicinity of each resonance. Provided this approximation is justified, the physics discussed above is also applicable in the single-photon regime. Indeed, in Fig.~\ref{fig4} the data at the lowest Rabi frequency corresponds to mean photon number $\braket{n^{}_{\mathrm{ph}}}\approx 1$, and clearly displays reduced loss at high sweep rates, suggesting that a treatment of independent TLSs is indeed relevant. A more thorough investigation of the single-photon regime will be performed elsewhere.

We note that the single-photon regime $\braket{n^{}_{\mathrm{ph}}}\ll 1$ corresponds to the problem of a qubit with energy splitting $\mathcal{E}^{}_{\mathrm{q}}$ coupled to a near-resonant TLS with energy splitting $\mathcal{E}^{}_{\mathrm{TLS}}$~\cite{Comment3}. Near resonance the relevant coupling is the transverse one, $\propto\sigma^{(\mathrm{q})}_{x}\sigma^{(\mathrm{TLS})}_{x}$, and within the subspace $\{\ket{0,e},\ket{1,g}\}$ ($\ket{0},\ket{1}$ and $\ket{g},\ket{e}$ being the qubit and the TLS ground and excited states, respectively) each resonance is again governed by the LZ dynamics. The above results thus suggest that by sweeping the bias energy of TLSs at a rate larger than their relaxation rate, but smaller than the qubit frequency $\omega^{}_{\mathrm{q}}=\mathcal{E}^{}_{\mathrm{q}}/\hbar$, one may dynamically decouple the qubit from sparse TLSs. Since this sweeping is slow compared to the time scale of the qubit dynamics, the qubit state remains unperturbed. This is in contrast to the saturation regime at strong resonant driving fields, where undesired qubit excitations are inevitable. 

\section*{Data Availability}
Data sets generated and analyzed during the current study are available from the corresponding author on request.

\section*{Acknowledgements}
This work was funded by the Deutsche Forschungsgesellschaft (DFG), grants LI2446/1-2 and SH 81//3-1), and by the Israel Science Foundation (ISF), grant No. 821/14. SM acknowledges support from the Minerva foundation. AB acknowledges support from the Helmholtz International Research School for Teratronics (HIRST) and the Landesgraduiertenf\"orderung-Karlsruhe (LGF). AVU acknowledges partial support from the Ministry of Education and Science of Russian Federation in the framework of the Increase Competitiveness Program of the National University of Science and Technology MISIS (Grant No. K2-2017-081).

\section*{Author Contributions}
S.M. developed the theoretical model and performed the calculations in collaboration with A.S. and M.S.; The samples were fabricated by H.S. and A.B.; Measurements were done by H.S. and J.L.; S.M. and J.L. wrote the paper with contributions from all authors.

\section*{Additional Information}
{\bf Supplementary information} accompanies this paper.\\
{\bf Competing interests:} The authors declare that there are no competing interests.


\begin{references}
\bibitem{Comment1} At the regime of low temperatures $k^{}_{\mathrm{B}}T\ll\hbar\omega$ considered here, the hyperbolic tangent factor is approximately unity, meaning that the thermal excitation of resonant TLSs is negligible.
\bibitem{Comment2} It is important to realize that the LZ transitions occur between photon number states that differ by one photon. This does not contradict the assumption of a classical resonator with mean photon number $\braket{n^{}_{\mathrm{ph}}}\gg 1$, such that the transition amplitudes are fixed for all TLSs and do not change between consecutive transitions, which allows us to treat TLSs independently.
\bibitem{Comment3} Note that a resonator is usually in a coherent state and not in a Fock state, and therefore its state cannot exactly be described as a superposition of two states. However, for mean photon number $\braket{n^{}_{\mathrm{ph}}}\ll 1$, the most important number states in the coherent superposition are the zero and one photon states, which are in correspondence to the ground and excited states of a qubit.
\end{references}
\end{document}


\title{Supplementary material for \\ Dynamical Decoupling of Quantum Two-Level Systems by Coherent Multiple Landau-Zener Transitions}
	
\author{Shlomi Matityahu}
\affiliation{Department of Physics, Ben-Gurion University of the Negev, Beer Sheva 84105, Israel}
\affiliation{Department of Physics, Nuclear Research Centre-Negev, Beer-Sheva 84190, Israel}
\affiliation{Institute of Nanotechnology, Karlsruhe Institute of Technology, D-76344 Eggenstein-Leopoldshafen, Germany}
\author{Hartmut Schmidt}
\affiliation{Physikalisches Institut, Karlsruhe Institute of Technology (KIT), 76131 Karlsruhe, Germany}
\author{Alexander Bilmes}
\affiliation{Physikalisches Institut, Karlsruhe Institute of Technology (KIT), 76131 Karlsruhe, Germany}
\author{Alexander Shnirman}
\affiliation{Institute of Nanotechnology, Karlsruhe Institute of Technology, D-76344 Eggenstein-Leopoldshafen, Germany}
\affiliation{Institut f\"ur Theorie der Kondensierten Materie, KIT, 76131 Karlsruhe, Germany}
\author{Georg Weiss}
\affiliation{Physikalisches Institut, Karlsruhe Institute of Technology (KIT), 76131 Karlsruhe, Germany}
\author{Alexey V. Ustinov}
\affiliation{Physikalisches Institut, Karlsruhe Institute of Technology (KIT), 76131 Karlsruhe, Germany}
\affiliation{Russian Quantum Center, National University of Science and Technology MISIS, Moscow 119049, Russia}
\author{Moshe Schechter}
\affiliation{Department of Physics, Ben-Gurion University of the Negev, Beer Sheva 84105, Israel}
\author{J\"urgen Lisenfeld}
\affiliation{Physikalisches Institut, Karlsruhe Institute of Technology (KIT), 76131 Karlsruhe, Germany}
	
\date{\today}

\keywords{} \maketitle

\section{Model} \label{Model}
We consider a single TLS with bias energy $\Delta(t)=\Delta(0)-2\mathbf{p}\cdot\mathbf{E}^{}_{\mathrm{bias}}(t)$ and tunneling energy $\Delta^{}_{0}$ in a classical single-mode resonator field $\mathbf{E}^{}_{\mathrm{res}}(t)=\mathbf{E}^{}_{\mathrm{ac}}\cos(\omega t)$. Here $\mathbf{p}$ is the TLS dipole moment and $\mathbf{E}^{}_{\mathrm{bias}}(t)\parallel\mathbf{E}^{}_{\mathrm{ac}}$ is a periodic field with period $T^{}_{\mathrm{sw}}\gg 2\pi/\omega$. The shortest periods in our experiment are $10\,$ns, which satisfy the latter condition for microwave resonators with resonance frequency $\omega\approx 2\pi\times 7\,$GHz. The Hamiltonian of the isolated TLS (without coupling to a dissipative bath) is 
\begin{align}
\label{eq:1}\mathcal{H}&=\frac{1}{2}\left(\Delta(t)\sigma^{}_{z}+\Delta^{}_{0}\sigma^{}_{x}\right)-\mathbf{p}\cdot\mathbf{E}^{}_{\mathrm{ac}}\cos(\omega t)\sigma^{}_{z}\nonumber\\
&=\frac{1}{2}\mathcal{E}(t)\left(\cos\theta(t)\sigma^{}_{z}+\sin\theta(t)\sigma^{}_{x}\right)-\mathbf{p}\cdot\mathbf{E}^{}_{\mathrm{ac}}\cos(\omega t)\sigma^{}_{z},
\end{align}
where $\mathcal{E}(t)=\sqrt{\Delta^{2}(t)+\Delta^{2}_{0}}$ is the TLS energy splitting, $\cos\theta(t)=\Delta(t)/\mathcal{E}(t)$ and $\sin\theta(t)=\Delta^{}_{0}/\mathcal{E}(t)$. We diagonalize the first part of the Hamiltonian by applying the transformation $u(t)=e^{-i\theta(t)\sigma^{}_{y}/2}$,
\begin{align}
\label{eq:2}&\mathcal{H}'=u\mathcal{H}u^{\dag}+i\hbar\dot{u}u^{\dag}=\frac{1}{2}\mathcal{E}(t)\sigma^{}_{z}-\mathbf{p}\cdot\mathbf{E}^{}_{\mathrm{ac}}\cos(\omega t)\left(\cos\theta\sigma^{}_{z}-\sin\theta\sigma^{}_{x}\right)-\frac{\hbar\dot{\theta}}{2}\sigma^{}_{y}.
\end{align}

Consider one period of the bias field $t\in\left[0,T^{}_{\mathrm{sw}}\right]$. A TLS with $\Delta^{}_{0}<\hbar\omega$ is swept through resonance at time $t^{}_{0}$ for which $\mathcal{E}(t^{}_{0})=\hbar\omega$ [Fig.~1b) in the main text]. Near this resonance the energy splitting can be expanded as~\cite{BAL13,KMS14} 
\begin{align}
\label{eq:3}&\mathcal{E}(t)=\sqrt{\Delta^{2}(t)+\Delta^{2}_{0}}\approx\hbar\omega+v(t-t^{}_{0}),
\end{align}
where 
\begin{align}
\label{eq:4}&v=\dot{\mathcal{E}}(t^{}_{0})=v^{}_{0}\cos\eta\sqrt{1-\left(\frac{\Delta^{}_{0}}{\hbar\omega}\right)^{2}},
\end{align}
with $\eta$ the angle between the TLS dipole moment and the electric fields and
\begin{align}
\label{eq:5}&v^{}_{0}=2p\dot{E}^{}_{\mathrm{bias}}(t^{}_{0}).
\end{align}
For the triangular bias field applied in our experiment [see Fig.~1a) in the main text], the sweep velocity is $|v^{}_{0}|=4pE^{}_{\mathrm{max}}/T^{}_{\mathrm{sw}}$. From the above definitions one has $\hbar|\dot{\theta}(t^{}_{0})|=|\Delta^{}_{0}v/(\Delta(t^{}_{0})\omega)|=(\Delta^{}_{0}/\hbar\omega)(v^{}_{0}/\omega)\cos\eta$. The maximum bias field rate in our experiment is $\dot{E}^{}_{\mathrm{bias}}=1.8\cdot10^{14}\,\mathrm{V}/(\mathrm{m}\cdot \mathrm{s})$, which yields $|\dot{\theta}(t^{}_{0})|/\omega<10^{-2}$ (assuming a typical value of $p=0.5\,$e$\AA$ for the TLS dipole moment). We can therefore safely neglect the last term in Eq.~(\ref{eq:2}). Moreover, near resonance ($\mathcal{E}\approx\hbar\omega$) and for $\Omega^{}_{\mathrm{R0}}\equiv pE^{}_{\mathrm{ac}}/\hbar\ll\omega$, the longitudinal coupling $\propto\sigma^{}_{z}$ is irrelevant compared to the transverse one $\propto\sigma^{}_{x}$. As a result, near each resonance the Hamiltonian~(\ref{eq:2}) can be reduced to
\begin{align}
\label{eq:6}&\mathcal{H}'\approx\frac{1}{2}\mathcal{E}(t)\sigma^{}_{z}+\hbar\Omega^{}_{\mathrm{R}}\cos(\omega t)\sigma^{}_{x},
\end{align}
where
\begin{align}
\label{eq:7}&\Omega^{}_{\mathrm{R}}=\Omega^{}_{\mathrm{R0}}\cos\eta\sin\theta\approx\Omega^{}_{\mathrm{R0}}\frac{\Delta^{}_{0}}{\hbar\omega}\cos\eta
\end{align}
is the TLS Rabi frequency. Finally, we transform to the rotating frame of reference by applying the transformation $u^{}_{\mathrm{R}}=e^{i\omega t\sigma^{}_{z}/2}$, 
\begin{align}
\label{eq:8}&\mathcal{H}'_{\mathrm{R}}=u^{}_{\mathrm{R}}\mathcal{H}'u^{\dag}_{\mathrm{R}}+i\hbar\dot{u}^{}_{\mathrm{R}}u^{\dag}_{\mathrm{R}}=\frac{1}{2}\left(\mathcal{E}(t)-\hbar\omega\right)\sigma^{}_{z}+\hbar\Omega^{}_{\mathrm{R}}\cos(\omega t)\left(\sigma^{}_{+}e^{i\omega t}+\sigma^{}_{-}e^{-i\omega t}\right),
\end{align}
where $\sigma^{}_{\pm}=(\sigma^{}_{x}\pm i\sigma^{}_{y})/2$. Under the conditions above, the rotating wave approximation can be invoked and the dynamics in the vicinity of each resonance is governed by the Landau-Zener (LZ) Hamiltonian
\begin{align}
\label{eq:9}&\mathcal{H}'_{\mathrm{R}}\approx\mathcal{H}^{}_{\mathrm{LZ}}=\frac{1}{2}v(t-t^{}_{0})\sigma^{}_{z}+\frac{1}{2}\hbar\Omega^{}_{\mathrm{R}}\sigma^{}_{x}.
\end{align}
Note that for the theory to apply, the period $T^{}_{\mathrm{sw}}$ of the bias field should be longer than the time interval for a single LZ transition, $t^{}_{\mathrm{LZ}}=\hbar\Omega^{}_{\mathrm{R}0}/|v^{}_{0}|$. This translates into the necessary condition $pE^{}_{\mathrm{max}}>\hbar\Omega^{}_{\mathrm{R}0}$, or $\xi^{}_{2}>\xi^{}_{1}$, where $\xi^{}_{1}\equiv2\Gamma^{}_{1}/(\pi\Omega^{}_{\mathrm{R}0})$ and $\xi^{}_{2}\equiv 8pE^{}_{\mathrm{max}}\Gamma^{}_{1}/(\pi\hbar\Omega^{2}_{\mathrm{R}0})$. This condition applies to all the experimental data presented in the paper. Note also that the results below assume non-resonant bias fields with $T^{}_{\mathrm{sw}}\gg 2\pi/\omega$ (applicable to all experimental data), as discussed above. In addition, the assumption that this bias field changes the energy splitting of the TLS adiabatically is valid provided that the adiabatic condition $|v^{}_{0}|/(\hbar\omega^{2})\ll 1$ holds. As discussed above, in our experiment $|v^{}_{0}|/(\hbar\omega^{2})<10^{-2}$, such that the adiabatic condition is satisfied.

\section{Liouville space} \label{Liouville}
We work in Liouville space, i.e.\ the linear space spanned by all linear operators acting on the Hilbert space. In a Hilbert space $\mathcal{H}^{}_{n}$ of dimension $n$, the elements are state vectors $\ket{\psi}$ spanned by a set of orthonormal basis states $\{\ket{i}\}^{n}_{i=1}$, namely $\ket{\psi}=\sum^{n}_{i=1}c^{}_{i}\ket{i}$ with $c^{}_{i}=\braket{i|\psi}$. An operator $\hat{O}$ acting on the state vectors can be represented as $\hat{O}=\sum^{n}_{i,j=1}O^{}_{ij}\ket{i}\bra{j}$ with $O^{}_{ij}=\braket{i|\hat{O}|j}$. Since any linear combination of these operators is also a linear operator, the set of linear operators $\ket{\hat{O}}$ acting over the Hilbert space with an inner product $\braket{\hat{O}^{}_{1}|\hat{O}^{}_{2}}=\mathrm{Tr}(\hat{O}^{\dag}_{1}O^{}_{2})$ forms a linear space known as the Liouville space. This space is spanned by the basis $\{\ket{i}\otimes\ket{j}\}^{n}_{i,j=1}$ and its dimension is $n^{2}$. Here the basis vector $\ket{i}\otimes\ket{j}$ corresponds to the Hilbert space operator $\ket{i}\bra{j}$. Linear operators in Liouville space acting on the elements $\ket{\hat{O}}$ are called superoperators and will be denoted as $\hat{\hat{L}}$.

Let us find the superoperator $\hat{\hat{L}}$ corresponding to the operator $\hat{A}\hat{O}\hat{B}$, i.e.\ the superoperator $\hat{\hat{L}}$ for which $\hat{\hat{L}}\ket{\hat{O}}=\ket{\hat{A}\hat{O}\hat{B}}$. This is obtained by transforming $\hat{A}\hat{O}\hat{B}=\sum^{n}_{i,j,k,l=1}A^{}_{ij}O^{}_{jk}B^{}_{kl}\ket{i}\bra{l}$ into
\begin{align}
\label{eq:10}\ket{\hat{A}\hat{O}\hat{B}}&=\sum^{n}_{i,j,k,l=1}A^{}_{ij}O^{}_{jk}B^{}_{kl}\ket{i}\otimes\ket{l}=\sum^{n}_{j,k=1}O^{}_{jk}\left(\hat{A}\ket{j}\otimes\hat{B}^{T}\ket{k}\right)\nonumber\\
&=\hat{A}\otimes\hat{B}^{T}\sum^{n}_{j,k=1}O^{}_{jk}\ket{j}\otimes\ket{k}=\hat{A}\otimes\hat{B}^{T}\ket{\hat{O}},
\end{align}
which shows that $\hat{\hat{L}}=\hat{A}\otimes\hat{B}^{T}$. For example, a unitary evolution of the density matrix, $\hat{\rho}\rightarrow\hat{\mathcal{U}}\hat{\rho}\,\hat{\mathcal{U}}^{\dag}$, is described in Liouville space as $\ket{\hat{\rho}}\rightarrow\hat{\mathcal{U}}\otimes\hat{\mathcal{U}}^{\ast}\ket{\hat{\rho}}$. 

Below and in the main text we omit the caret symbol from Hilbert space operators and Liouville space vectors and superoperators. Accordingly, the $2\times 2$ density matrix operator in Hilbert space will be denoted as $\rho$, and its representation as a 4-vector in Liouville space will be denoted as $\ket{\rho}=\left(\rho^{}_{00},\rho^{}_{01},\rho^{}_{10},\rho^{}_{11}\right)^{T}$. The distinction between Hilbert space operators and Liuoville space superoperators should be clear by the context.

\section{Full-counting statistics for photon absorption by a TLS} \label{FCS}
\begin{figure}[htb!]
	\includegraphics[width=0.7\columnwidth,height=7.5cm]{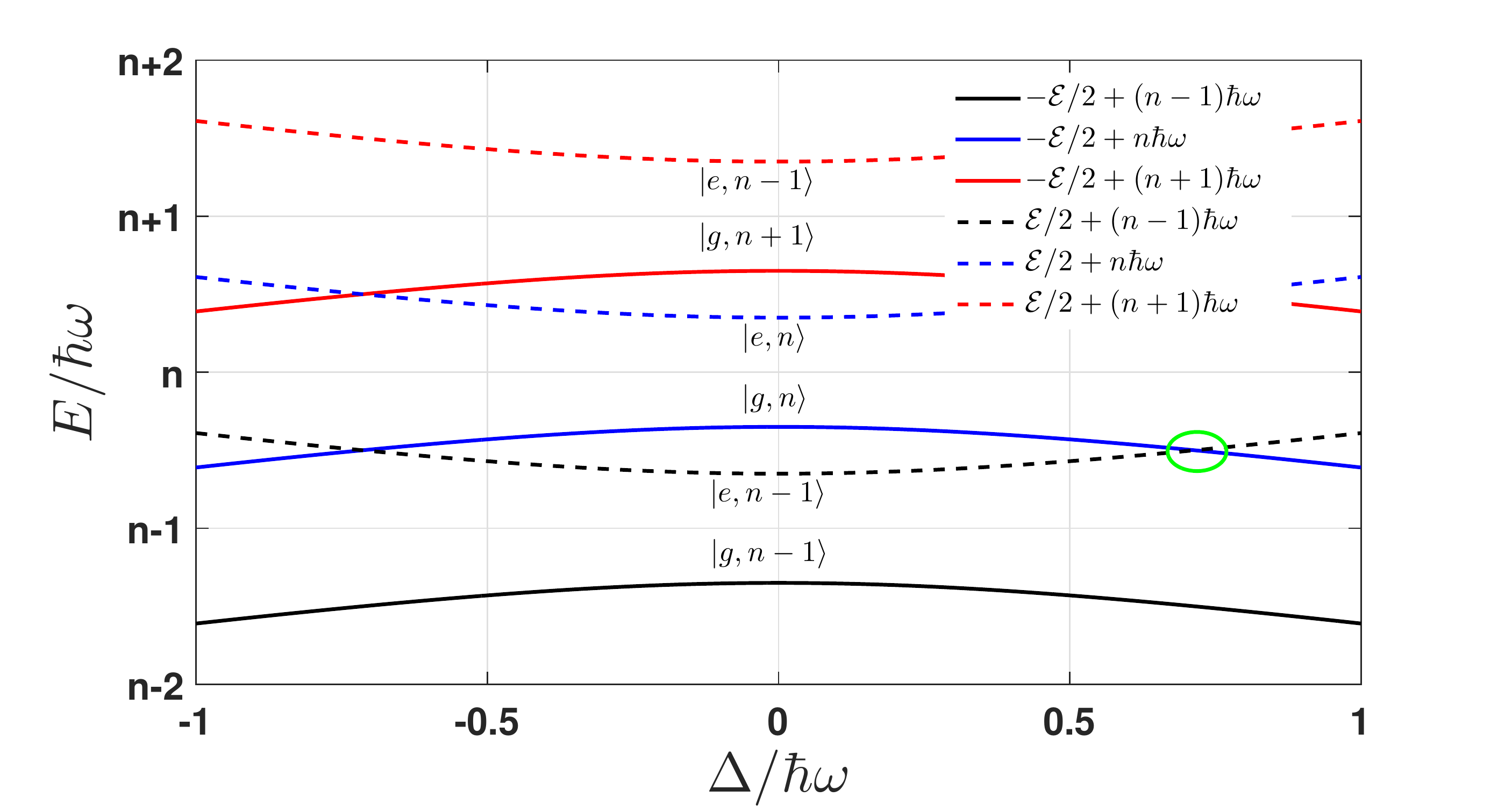}
	\caption{Energy levels of the non-interacting TLS-field Hamiltonian $\mathcal{H}^{}_{0}=\left(\mathcal{E}/2\right)\sigma^{}_{z}+\hbar\omega a^{\dag}a$ as a function of the bias energy $\Delta$, for $\Delta^{}_{0}=0.7$ (in units of $\hbar\omega$). The green circle shows the resonance between the states $\ket{g,n}$ and $\ket{e,n-1}$. Subsequent resonances described in Fig.~1 in the main text correspond to crossings of this resonance in opposite directions.} 
	\label{fig1}
\end{figure}

To calculate the full counting statistics of the photon number absorbed by a TLS after time $t=NT^{}_{\mathrm{sw}}$, we notice that in a fully quantized description of matter and light, the LZ transitions described above are transitions between the diabatic states $\ket{g,n}$ and $\ket{e,n-1}$ of the combined TLS-field system ($\ket{g}$ and $\ket{e}$ being the ground and excited states of the TLS, and $\ket{n}$ being a photon number state). The resonance in this fully quantized picture is shown in Fig.~\ref{fig1}, which plots the energy levels of the non-interacting TLS-field Hamiltonian $\mathcal{H}^{}_{0}=\left(\mathcal{E}/2\right)\sigma^{}_{z}+\hbar\omega a^{\dag}a$ ($a^{\dag}$ and $a$ are the photon creation and annihilation operators) as a function of the bias energy $\Delta$. Below we consider the periodic back and forth crossings of one of these resonances. 

We insert a counting field $k$ into the evolution operator describing a single LZ transition~\cite{SSN10} governed by the Hamiltonian~(\ref{eq:9}),
\begin{align}
\label{eq:11}&
\mathcal{U}^{}_{\mathrm{LZ}}(k)=
\begin{pmatrix}
\sqrt{P} & e^{i\frac{k}{2}}e^{-i\psi}\sqrt{1-P} \\
-e^{-i\frac{k}{2}}e^{i\psi}\sqrt{1-P} & \sqrt{P}
\end{pmatrix},
\end{align}
where $P=e^{-1/\xi}$ is the non-adiabatic transition probability with $\xi=2|v|/(\pi\hbar\Omega^{2}_{\mathrm{R}})$, and $\psi=\pi/4+\arg\Gamma(1-i/(2\pi\xi))-\left[\ln(2\pi\xi)+1\right]/(2\pi\xi)$ is the so-called Stokes phase (here $\Gamma$ is the gamma function)~\cite{SSN10}. In the presence of the counting field $k$, the appropriate superoperator corresponding to the evolution of the density matrix $\ket{\rho}$ is
\begin{align}
\label{eq:12}&
U^{}_{\mathrm{LZ}}=\mathcal{U}^{}_{\mathrm{LZ}}(k)\otimes\left[\mathcal{U}^{}_{\mathrm{LZ}}(-k)\right]^{\ast},
\end{align}
since the counting field changes sign on the two Keldysh contours, i.e.\ the evolution of the density matrix in Hilbert space is $\hat{\rho}\rightarrow\hat{\mathcal{U}}^{}_{\mathrm{LZ}}(k)\hat{\rho}\,\hat{\mathcal{U}}^{\dag}_{\mathrm{LZ}}(-k)$ (see, e.g., Eq.~(11) in Ref.~\cite{LLS96}). It should be noted that the evolution operator~(\ref{eq:11}) corresponds to the Hamiltonian~(\ref{eq:9}) with positive sign of the sweep velocity $v$~\cite{SSN10}. Without a counting field, the evolution operator corresponding to the Hamiltonian~(\ref{eq:9}) with negative sign of $v$ is obtained by swapping the off-diagonal elements~\cite{SSN10,Comment}. Since the exponential terms corresponding to the counting field are not affected by the sign of $v$, the appropriate transformation for the evolution operator~(\ref{eq:11}) is $\psi\rightarrow\pi-\psi$.    

In between two successive transitions, the TLS is out of resonance for a time interval $t$, and its interaction with the resonator ac field is not important. The TLS dynamics within this time interval is described by the Lindblad master equation,
\begin{align}
\label{eq:13}&
\dot{\rho}=-\frac{i}{\hbar}\left[\mathcal{H}^{}_{\mathrm{TLS}},\rho\right]+\sum^{}_{i=\pm}\Gamma^{}_{i}\left(L^{}_{i}\rho L^{\dag}_{i}-\frac{1}{2}\{L^{\dag}_{i}L^{}_{i},\rho\}\right),
\end{align}  
where $\mathcal{H}^{}_{\mathrm{TLS}}(t)=(\mathcal{E}(t)/2)\sigma^{}_{z}$, $L^{}_{\pm}=\sigma^{}_{\pm}=(\sigma^{}_{x}\pm i\sigma^{}_{y})/2$ and $\Gamma^{}_{+}=\Gamma^{}_{\uparrow}$, $\Gamma^{}_{-}=\Gamma^{}_{\downarrow}$  are the transition rates between the TLS eigenstates. For simplicity, we assume no pure dephasing, such that the decoherence rate is $\Gamma^{}_{2}=\Gamma^{}_{1}/2$, where $\Gamma^{}_{1}=\Gamma^{}_{\uparrow}+\Gamma^{}_{\downarrow}$ is the relaxation rate. The components of this master equation read
\begin{align}
\label{eq:14}&
\dot{\rho}^{}_{00}=-\Gamma^{}_{\downarrow}\rho^{}_{00}+\Gamma^{}_{\uparrow}\rho^{}_{11},\nonumber\\
&\dot{\rho}^{}_{01}=-\left(i\frac{\mathcal{E}(t)}{\hbar}-\Gamma^{}_{2}\right)\rho^{}_{01},\nonumber\\
&\dot{\rho}^{}_{10}=\left(i\frac{\mathcal{E}(t)}{\hbar}-\Gamma^{}_{2}\right)\rho^{}_{10},\nonumber\\
&\dot{\rho}^{}_{11}=\Gamma^{}_{\downarrow}\rho^{}_{00}-\Gamma^{}_{\uparrow}\rho^{}_{11}.
\end{align}
Taking into account that $\mathrm{Tr}\left[\rho(t)\right]=\rho^{}_{00}(t)+\rho^{}_{11}(t)=1$ at any time $t$, the solution of these equations is $\ket{\rho(t)}=U(t)\ket{\rho(0)}$, with the superoperator 
\begin{align}
\label{eq:15}&
U(t)=
\begin{pmatrix}
\frac{\Gamma^{}_{\uparrow}}{\Gamma^{}_{1}}+\frac{\Gamma^{}_{\downarrow}}{\Gamma^{}_{1}}e^{-\Gamma^{}_{1}t} & 0 & 0 & \frac{\Gamma^{}_{\uparrow}}{\Gamma^{}_{1}}\left(1-e^{-\Gamma^{}_{1}t}\right) \\
0 & e^{i\phi(t)-\Gamma^{}_{2}t} & 0 & 0 \\
0 & 0 & e^{-i\phi(t)-\Gamma^{}_{2}t} & 0 \\
\frac{\Gamma^{}_{\downarrow}}{\Gamma^{}_{1}}\left(1-e^{-\Gamma^{}_{1}t}\right) & 0 & 0 & \frac{\Gamma^{}_{\downarrow}}{\Gamma^{}_{1}}+\frac{\Gamma^{}_{\uparrow}}{\Gamma^{}_{1}}e^{-\Gamma^{}_{1}t}
\end{pmatrix},
\end{align}
where $\phi(t)=\frac{1}{\hbar}\int^{t}_{0}\mathcal{E}(t')dt'$. The evolution of the density matrix after one period of the bias field is given by $\ket{\rho(T^{}_{\mathrm{sw}},k)}=U^{}_{\mathrm{sw}}(k)\ket{\rho(0)}$, where $U^{}_{\mathrm{sw}}(k)=U^{}_{\mathrm{LZ}}(\pi-\psi,k)U(t^{}_{2})U^{}_{\mathrm{LZ}}(\psi,k)U(t^{}_{1})$ with $T^{}_{\mathrm{sw}}=t^{}_{1}+t^{}_{2}$. Note that the matrices describing the LZ transitions differ by the Stokes phase, as discussed above, since two subsequent transitions have different sign of velocity [see Fig.~1b) in the main text]. The evolution after time $t=NT^{}_{\mathrm{sw}}$ is then
\begin{align}
\label{eq:16}&
\ket{\rho(t,k)}=U^{N}_{\mathrm{sw}}(k)\ket{\rho(0)}.
\end{align}

Letting $p(t,n)$ be the probability of dissipating $n$ photons at time $t$, we define the generating function
\begin{align}
\label{eq:17}&
\chi(t,k)=\sum^{\infty}_{n=0}e^{ikn}p(t,n),
\end{align}
such that the moments of $p(t,n)$ are given by
\begin{align}
\label{eq:18}&
\braket{N^{m}_{\mathrm{ph}}(t)}=\sum^{\infty}_{n=0}n^{m}p(t,n)=(-i)^{m}\frac{\partial^{m}\chi(t,k)}{\partial k^{m}}\bigg|^{}_{k=0}.
\end{align}
Within the full counting statistics formalism, the generating function is obtained by taking the partial trace over the TLS state, 
\begin{align}
\label{eq:19}
\chi(t,k)&=\mathrm{Tr}\left[\ket{\rho(t,k)}\right]=\mathrm{Tr}\left[U^{N}_{\mathrm{sw}}(k)\ket{\rho(0)}\right],
\end{align}
where the trace operation is defined as $\mathrm{Tr}\left[\ket{\rho}\right]\equiv\rho^{}_{00}+\rho^{}_{11}$. The quantity we are interested in is the average number of photons absorbed by the TLS during time $t=NT^{}_{\mathrm{sw}}$,
\begin{align}
\label{eq:20}
\braket{N^{}_{\mathrm{ph}}(t)}&=-i\frac{\partial\chi(t,k)}{\partial k}\bigg|^{}_{k=0}=-i\frac{d}{dk}\mathrm{Tr}\left[U^{N}_{\mathrm{sw}}(k)\ket{\rho(0)}\right]\bigg|^{}_{k=0}\nonumber\\
&=-i\mathrm{Tr}\left[\frac{d}{dk}U^{N}_{\mathrm{sw}}(k)\bigg|^{}_{k=0}\ket{\rho(0)}\right].
\end{align}

To calculate this quantity, we write the superoparator $U^{}_{\mathrm{sw}}(k)$ in terms of its eigendecomposition, $U^{}_{\mathrm{sw}}(k)=V^{}_{k}\Lambda^{}_{k}G^{}_{k}$, where $\Lambda(k)=\mathrm{diag}\{\lambda^{}_{1}(k),\lambda^{}_{2}(k),\lambda^{}_{3}(k),\lambda^{}_{4}(k)\}$ is the diagonal matrix of eigenvalues $\{\lambda^{}_{j}(k)\}^{4}_{j=1}$, and $V^{}_{k}$, $G^{}_{k}=V^{-1}_{k}$ are the matrices whose columns and rows are the corresponding right and left eigenvectors, respectively. We obtain,
\begin{align}
\label{eq:21}
\braket{N^{}_{\mathrm{ph}}(t)}&=-i\mathrm{Tr}\left[\frac{d}{dk}\left(V^{}_{k}\Lambda^{N}_{k}G^{}_{k}\right)\bigg|^{}_{k=0}\ket{\rho(0)}\right]\nonumber\\
&=-i\mathrm{Tr}\bigg[\big(V'^{}_{0}\Lambda^{N}_{0}G^{}_{0}+NV^{}_{0}\Lambda'^{}_{0}\Lambda^{N-1}_{0}G^{}_{0}+V^{}_{0}\Lambda^{N}_{0}G'^{}_{0}\big)\ket{\rho(0)}\bigg].
\end{align}
In the limit $N\rightarrow\infty$ only the middle term of Eq.~(\ref{eq:21}) will contribute to the photon absorption rate per TLS, $\gamma^{}_{\mathrm{abs}}=\lim_{t\rightarrow\infty}\braket{N^{}_{\mathrm{ph}}(t)}/t$, and therefore
\begin{align}
\label{eq:22}&
\gamma^{}_{\mathrm{abs}}=\lim^{}_{t\rightarrow\infty}\frac{\braket{N^{}_{\mathrm{ph}}(t)}}{t}=\lim^{}_{N\rightarrow\infty}\frac{\braket{N^{}_{\mathrm{ph}}(t=NT^{}_{\mathrm{sw}})}}{NT^{}_{\mathrm{sw}}}=-\frac{i}{T^{}_{\mathrm{sw}}}\lim^{}_{N\rightarrow\infty}\mathrm{Tr}\left[V^{}_{0}\Lambda'^{}_{0}\Lambda^{N-1}_{0}G^{}_{0}\ket{\rho(0)}\right].
\end{align}
To calculate Eq.~(\ref{eq:22}), one needs to diagonalize a $4\times 4$ matrix. Further analytical progress can be achieved by inserting the identity matrix $G^{}_{0}V^{}_{0}$ into Eq.~(\ref{eq:20}),
\begin{align}
\label{eq:23}&
\gamma^{}_{\mathrm{abs}}=-\frac{i}{T^{}_{\mathrm{sw}}}\lim^{}_{N\rightarrow\infty}\mathrm{Tr}\left[V^{}_{0}\Lambda'^{}_{0}G^{}_{0}V^{}_{0}\Lambda^{N-1}_{0}G^{}_{0}\ket{\rho(0)}\right].
\end{align}
Since for $k=0$ we have $\mathrm{Tr}\left[\ket{\rho(t,k=0)}\right]=1$ for any time $t$, Eq.~(\ref{eq:16}) implies that $\bra{g^{}_{1}}=\left(1,0,0,1\right)$ is a left eigenvector of $U^{}_{\mathrm{sw}}(0)$ with eigenvalue $\lambda^{}_{1}(0)=1$. For a stationary solution, the other eigenvalues satisfy $|\lambda^{}_{j}(0)|<1$ for $j=2,3,4$. Thus, in the limit $N\rightarrow\infty$, $V^{}_{0}\Lambda^{N-1}_{0}G^{}_{0}$ reduces to $\ket{v^{}_{1}}\bra{g^{}_{1}}$, where $\ket{v^{}_{1}}$ is the right eigenvector of $U^{}_{\mathrm{sw}}(k=0)$ corresponding to the eigenvalue $\lambda^{}_{1}(k=0)=1$. Hence,
\begin{align}
\label{eq:24}
\gamma^{}_{\mathrm{abs}}&=-\frac{i}{T^{}_{\mathrm{sw}}}\mathrm{Tr}\left[V^{}_{0}\Lambda'^{}_{0}G^{}_{0}\ket{v^{}_{1}}\braket{g^{}_{1}|\rho(0)}\right]=-\frac{i}{T^{}_{\mathrm{sw}}}\mathrm{Tr}\left[V^{}_{0}\Lambda'^{}_{0}G^{}_{0}\ket{v^{}_{1}}\right]\nonumber\\
&=-\frac{i}{T^{}_{\mathrm{sw}}}\braket{g^{}_{1}|V^{}_{0}\Lambda'^{}_{0}G^{}_{0}|v^{}_{1}}.
\end{align}
Finally, we substitute $\Lambda^{}_{0}=G^{}_{0}U^{}_{\mathrm{sw}}(0)V^{}_{0}$ to obtain
\begin{align}
\label{eq:25}
\gamma^{}_{\mathrm{abs}}&=-\frac{i}{T^{}_{\mathrm{sw}}}\braket{g^{}_{1}|V^{}_{0}\Lambda'^{}_{0}G^{}_{0}|v^{}_{1}}=-\frac{i}{T^{}_{\mathrm{sw}}}\braket{g^{}_{1}|V^{}_{0}G'^{}_{0}U^{}_{\mathrm{sw}}(0)+U'^{}_{\mathrm{sw}}(0)+U^{}_{\mathrm{sw}}(0)V'^{}_{0}G^{}_{0}|v^{}_{1}}\nonumber\\
&=-\frac{i}{T^{}_{\mathrm{sw}}}\braket{g^{}_{1}|\frac{dU^{}_{\mathrm{sw}}}{dk}\bigg|^{}_{k=0}|v^{}_{1}},
\end{align}
where in the last step we have used the fact that $\ket{v^{}_{1}}$ and $\bra{g^{}_{1}}$ are right and left eigenvectors of $U^{}_{\mathrm{sw}}(0)$ corresponding to an eigenvalue $\lambda^{}_{1}(0)=1$, and $V^{}_{0}G'^{}_{0}+V'^{}_{0}G^{}_{0}=\frac{d}{dk}\left(V^{}_{k}G^{}_{k}\right)|^{}_{k=0}=0$.

\section{TLS photon absorption rate and dielectric loss tangent} \label{absorption rate}
The evaluation of Eq.~(\ref{eq:25}) involves a calculation of the derivative $dU^{}_{\mathrm{sw}}/dk|^{}_{k=0}$ and the right eigenvector $\ket{v^{}_{1}}$ of $U^{}_{\mathrm{sw}}(0)$ corresponding to an eigenvalue $\lambda^{}_{1}(0)=1$, and can therefore be done analytically. At low temperatures $k^{}_{\mathrm{B}}T\ll\hbar\omega$ one has $\Gamma^{}_{\uparrow}\ll\Gamma^{}_{\downarrow}$ (and thus $\Gamma^{}_{1}\approx\Gamma^{}_{\downarrow}$), and we end up with $\gamma^{}_{\mathrm{abs}}=a/b$, where
\begin{align}
\label{eq:26}&
a=\left(P-1\right)\bigg\{4\left(1-P\right)\sinh\left(\frac{\Gamma^{}_{1}t^{}_{2}}{2}\right)\sinh\left(\frac{\Gamma^{}_{1}t^{}_{2}}{2}\right)\cos\left(\tilde{\phi}^{}_{1}-\tilde{\phi}^{}_{2}\right)+\nonumber\\
&2P\sinh\left(\frac{\Gamma^{}_{1}T^{}_{\mathrm{sw}}}{2}\right)\left[\sinh\left(\frac{\Gamma^{}_{1}t^{}_{2}}{2}\right)\cos\tilde{\phi}^{}_{1}+\sinh\left(\frac{\Gamma^{}_{1}t^{}_{1}}{2}\right)\cos\tilde{\phi}^{}_{2}\right]+2P-1-\cosh\left(\Gamma^{}_{1}T^{}_{\mathrm{sw}}\right)+\nonumber\\
&\left[\cosh\left(\Gamma^{}_{1}t^{}_{1}\right)+\cosh\left(\Gamma^{}_{1}t^{}_{2}\right)\right]\left(1-P\right)\bigg\},\nonumber\\
&b=2T^{}_{\mathrm{sw}}\bigg\{\frac{1}{4}\sinh\left(\Gamma^{}_{1}T^{}_{\mathrm{sw}}\right)+\frac{1}{2}\sinh\left(\frac{\Gamma^{}_{1}T^{}_{\mathrm{sw}}}{2}\right)\cos\left(\tilde{\phi}^{}_{1}-\tilde{\phi}^{}_{2}\right)\left(2P-1\right)+\nonumber\\
&P\left(P-1\right)\left[\sinh\left(\frac{\Gamma^{}_{1}t^{}_{1}}{2}\right)\cos\tilde{\phi}^{}_{1}+\sinh\left(\frac{\Gamma^{}_{1}t^{}_{2}}{2}\right)\cos\tilde{\phi}^{}_{2}\right]-\nonumber\\
&\sinh\left(\frac{\Gamma^{}_{1}T^{}_{\mathrm{sw}}}{2}\right)P^2\left(\cos\tilde{\phi}^{}_{1}+\cos\tilde{\phi}^{}_{1}\right)\bigg\},
\end{align}
with $\tilde{\phi}^{}_{1}=\phi^{}_{1}-2\psi$ and $\tilde{\phi}^{}_{2}=\phi^{}_{2}+2\psi$. Note that the absorption rate $\gamma^{}_{\mathrm{abs}}$ depends on the parameters $\Delta(0)$, $\Delta^{}_{0}$ and $\eta$ of each TLS, via its dependence on $P$, $\Gamma^{}_{1}$, $t^{}_{1}$, $\phi^{}_{1}$, $\phi^{}_{2}$ and $\psi$.

The expression relating the dielectric loss tangent to the total photon absorption rate per unit volume, $\Gamma^{}_{\mathrm{abs}}$, is obtained by comparing the equivalent expressions for the power dissipation energy, $P^{}_{\mathrm{dis}}=-\hbar\omega\Gamma^{}_{\mathrm{abs}}=-\frac{1}{2}\omega\epsilon''E^{2}_{\mathrm{ac}}$. This gives
\begin{align}
\label{eq:27}&
\tan\delta=\frac{\epsilon''}{\epsilon'}=\frac{2\hbar\Gamma^{}_{\mathrm{abs}}}{\epsilon E^{2}_{\mathrm{ac}}}=\frac{2p^{2}\Gamma^{}_{\mathrm{abs}}}{\epsilon\hbar\Omega^{2}_{\mathrm{R}0}}.
\end{align}
The total absorption rate per unit volume is obtained by averaging over the distribution of TLSs and the orientation of their dipole moments, as follows. For a given value of $\Delta^{}_{0}<\hbar\omega$ and $\eta$, the TLSs that are swept into resonance are those with bias energy $\Delta(0)$ in the window $2pE^{}_{\mathrm{max}}|\cos\eta|$ around $\sqrt{\left(\hbar\omega\right)^{2}-\Delta^{2}_{0}}$. The total photon absorption rate per unit volume is thus
\begin{align}
\label{eq:28}
\Gamma^{}_{\mathrm{abs}}&=\int^{\pi}_{0}d\eta\frac{\sin\eta}{2}\int d^{2}n^{}_{\mathrm{TLS}}\gamma^{}_{\mathrm{abs}}(\Delta(0),\Delta^{}_{0},\eta)\nonumber\\
&=2pE^{}_{\mathrm{max}}\int^{\hbar\omega}_{0}d\Delta^{}_{0}\frac{P^{}_{0}}{\Delta^{}_{0}}\int^{\pi}_{0}d\eta\frac{\sin\eta}{2}|\cos\eta|\gamma^{}_{\mathrm{abs}}(\Delta^{}_{0},\eta),
\end{align}
where $d^{2}n^{}_{\mathrm{TLS}}=\left(P^{}_{0}/\Delta^{}_{0}\right)d\Delta d\Delta^{}_{0}$ is the number of TLSs per unit volume with bias and tunneling energies in an element $d\Delta d\Delta^{}_{0}$ around $(\Delta,\Delta^{}_{0})$. In the last step of Eq.~(\ref{eq:28}) we assumed that $\gamma^{}_{\mathrm{abs}}$ is independent of $\Delta(0)$ so that the integral over the bias energy becomes trivial. 

We first consider the incoherent limit $\Gamma^{}_{1}T^{}_{\mathrm{sw}}\gg 1$, already discussed in Refs.~\cite{BAL13,KMS14}. In terms of $\xi$, this limit can be expressed as $\xi\ll\xi^{}_{2}$ with $\xi^{}_{2}\equiv 8pE^{}_{\mathrm{max}}\Gamma^{}_{1}/(\pi\hbar\Omega^{2}_{\mathrm{R}0})$. In this limit we have $a\approx\left(1-P\right)\cosh\left(\Gamma^{}_{1}T^{}_{\mathrm{sw}}\right)\approx\left(1-P\right)e^{\Gamma^{}_{1}T^{}_{\mathrm{sw}}}/2$ and $b\approx T^{}_{\mathrm{sw}}\sinh\left(\Gamma^{}_{1}T^{}_{\mathrm{sw}}\right)/2\approx T^{}_{\mathrm{sw}}e^{\Gamma^{}_{1}T^{}_{\mathrm{sw}}}/4$ in the leading order. Thus,
\begin{align}
\label{eq:29}&
\gamma^{}_{\mathrm{abs}}\approx\frac{2\left(1-P\right)}{T^{}_{\mathrm{sw}}},
\end{align}
which is independent of the phases $\phi^{}_{1}$ and $\phi^{}_{2}$ and the time intervals $t^{}_{1}$ and $t^{}_{2}$, and thus independent of $\Delta(0)$. Using Eq.~(\ref{eq:28}), the loss tangent~(\ref{eq:27}) takes the form
\begin{align}
\label{eq:30}\tan\delta&\approx \frac{2P^{}_{0}p^{2}v^{}_{0}}{\epsilon\hbar\Omega^{2}_{\mathrm{R}0}}\int^{\hbar\omega}_{0}\frac{d\Delta^{}_{0}}{\Delta^{}_{0}}\int^{\pi}_{0}d\eta\frac{\sin\eta}{2}|\cos\eta|\left(1-e^{-\pi\hbar\Omega^{2}_{\mathrm{R}}/2|v|}\right)\nonumber\\
&=\frac{2P^{}_{0}p^{2}}{\epsilon\hbar}\int^{\hbar\omega}_{0}\frac{d\Delta^{}_{0}}{\Delta^{}_{0}}\frac{\left(\Delta^{}_{0}/\hbar\omega\right)^{2}}{\sqrt{1-(\Delta^{}_{0}/\hbar\omega)^{2}}}\int^{\pi}_{0}d\eta\frac{\sin\eta}{2}\cos^{2}\eta\,\frac{|v|}{\Omega^{2}_{\mathrm{R}}}\left(1-e^{-\pi\hbar\Omega^{2}_{\mathrm{R}}/2|v|}\right),
\end{align}
where in the last step we have used Eqs.~(\ref{eq:4}) and~(\ref{eq:7}). Equation~(\ref{eq:30}) is analogous to Eq.~(6) of Ref.~\onlinecite{BAL13}, and reduces in the non-adiabatic limit $|v|\gg\hbar\Omega^{2}_{\mathrm{R}}$ to the intrinsic loss tangent $\tan\delta^{}_{0}=\pi P^{}_{0}p^{2}/(3\epsilon)$ obtained at low powers. This is readily observed by expanding the exponent in Eq.~(\ref{eq:30}) to first order,
\begin{align}
\label{eq:31}\tan\delta&\approx\frac{\pi P^{}_{0}p^{2}}{2\epsilon}\int^{\hbar\omega}_{0}\frac{d\Delta^{}_{0}}{\Delta^{}_{0}}\frac{\left(\Delta^{}_{0}/\hbar\omega\right)^{2}}{\sqrt{1-(\Delta^{}_{0}/\hbar\omega)^{2}}}\int^{\pi}_{0}d\eta\sin\eta\cos^{2}\eta=\frac{\pi P^{}_{0}p^{2}}{3\epsilon}.
\end{align}

An analytical or numerical evaluation of the loss tangent in the coherent regime $\Gamma^{}_{1}T^{}_{\mathrm{sw}}\ll 1$ using the full dependence of Eq.~(\ref{eq:26}) on the parameters $\Delta(0)$, $\Delta^{}_{0}$ and $\eta$ is quite complicated. One has to calculate the expression for the phases $\phi^{}_{1}$ and $\phi^{}_{2}$, as well as the time $t^{}_{1}$, in terms of these quantities, which produces a complicated dependence of $\gamma^{}_{\mathrm{abs}}$ on $\Delta(0)$, $\Delta^{}_{0}$ and $\eta$. As described by the first equality in Eq.~(\ref{eq:28}), one then has to average $\gamma^{}_{\mathrm{abs}}(\Delta(0),\Delta^{}_{0},\eta)$ over $\Delta(0)$ in the window $2pE^{}_{\mathrm{max}}|\cos\eta|$ around $\sqrt{\left(\hbar\omega\right)^{2}-\Delta^{2}_{0}}$ and then to average over the distribution of tunneling energies $\Delta^{}_{0}$ and orientation $\eta$ of the dipole moment with respect to the field. In order to fit the experimental data to theory, one should repeat this procedure with two fitting parameters, the dipole moment $p$ and the maximum relaxation rate of TLSs. To obtain quantitative comparison, one should also note the following issues. First, our analytical expression for $\gamma^{}_{\mathrm{abs}}$ in terms of the phases $\phi^{}_{1}$ and $\phi^{}_{2}$ assumes that a TLS undergoes two LZ transitions in a single period of the bias field, corresponding to crossings of one of the resonances shown in Fig.~\ref{fig1} in opposite directions. However, some TLSs undergo four transitions, corresponding to two crossings of each of the two resonances shown in Fig.~\ref{fig1}. This is not expected to change the results qualitatively, but can yield non-negligible quantitative differences. Second, the theory assumes a fixed amplitude $E^{}_{\mathrm{ac}}$ of the resonator field (and thus of the Rabi frequency $\Omega^{}_{\mathrm{R}0}=pE^{}_{\mathrm{ac}}$), such that the value of the parameter $\xi^{}_{2}$ is fixed in each theoretical curve shown in Fig.~2 in the main text. On the other hand, each dataset shown in Fig.~4 in the main text is taken at a fixed input power and a fixed amplitude $E^{}_{\mathrm{max}}$ of the bias field. The resonator field amplitude $E^{}_{\mathrm{ac}}$ depends on the loss of the resonator and is thus not fixed for each of these datasets. As a result, the value of $\xi^{}_{2}$ is not constant for each dataset shown in Fig.~4 in the main text, and the specified value of $\xi^{}_{2}$ is a rough estimate based on the value of $\Omega^{}_{\mathrm{R}0}$ at $\xi=0$. A quantitative comparison of experiment and theory would therefore require a self-consistent calculation of $\Omega^{}_{\mathrm{R}0}$ and $\tan\delta$. For these reasons, a direct fit of the data to theory is beyond the scope of this paper.
	
Instead, we choose to concentrate on the main effect of the ensemble of TLSs relevant to the interference over multiple coherent transitions, which is the distribution of the phases $\phi^{}_{1}$ and $\phi^{}_{2}$. We thus neglect the distribution of $\Delta^{}_{0}$, $p$ and $\eta$ in all other quantities, such as the sweep rate, the Rabi frequency, the relaxation rate and the stokes phase, and set $t^{}_{1}=t^{}_{2}=T^{}_{\mathrm{sw}}/2$ in Eq.~(\ref{eq:26}) (the qualitative results are not sensitive to the latter choice). In terms of the number of coherent transitions, $M=\left(\Gamma^{}_{1}T^{}_{\mathrm{sw}}\right)^{-1}=\xi/\xi^{}_{2}$, the numerator and denominator of the absorption rate $\gamma^{}_{\mathrm{abs}}=a/b$ [Eq.~(\ref{eq:26})] then become
\begin{align}
\label{eq:32}
&a=(P-1)\bigg\{4\sinh^{2}\left(\frac{1}{4M}\right)(1-P)\cos\left(\tilde{\phi}^{}_{1}-\tilde{\phi}^{}_{2}\right)+2P-1-\cosh\left(\frac{1}{M}\right)\nonumber\\
&\,\,\,\quad+2\sinh\left(\frac{1}{4M}\right)\sinh\left(\frac{1}{2M}\right)P\left(\cos\tilde{\phi}^{}_{1}+\cos\tilde{\phi}^{}_{2}\right)+2\cosh\left(\frac{1}{2M}\right)\left(1-P\right)\bigg\},\nonumber\\
&b=\frac{2}{\Gamma^{}_{1}M}\bigg\{P\left[\sinh\left(\frac{1}{4M}\right)(P-1)\left(\cos\tilde{\phi}^{}_{1}+\cos\tilde{\phi}^{}_{2}\right)-\sinh\left(\frac{1}{2M}\right)P\cos\tilde{\phi}^{}_{1}\cos\tilde{\phi}^{}_{2}\right]\nonumber\\
&\,\,\,\quad+\frac{1}{2}\sinh\left(\frac{1}{2M}\right)(2P-1)\cos(\tilde{\phi}^{}_{1}-\tilde{\phi}^{}_{2})+\frac{1}{4}\sinh\left(\frac{1}{M}\right)\bigg\}.
\end{align}
The absorption rate $\gamma^{}_{\mathrm{abs}}=a/b$ is then a function of $\xi$, $\xi^{}_{2}$, $\tilde{\phi}^{}_{1}$ and $\tilde{\phi}^{}_{2}$. Figure 2 in the main text shows the loss tangent obtained by averaging the absorption rate over a homogeneous distribution of the phases $\phi^{}_{1}$ and $\phi^{}_{2}$ (and thus of $\tilde{\phi}^{}_{1}$ and $\tilde{\phi}^{}_{2}$),
\begin{align}
\label{eq:33}
&\braket{\gamma^{}_{\mathrm{abs}}}(\xi,\xi^{}_{2})=\frac{1}{(2\pi)^{2}}\int^{2\pi}_{0}\int^{2\pi}_{0}\gamma^{}_{\mathrm{abs}}(\xi,\xi^{}_{2},\tilde{\phi}^{}_{1},\tilde{\phi}^{}_{2})d\tilde{\phi}^{}_{1}d\tilde{\phi}^{}_{2}.
\end{align}

To understand the behavior of this average, we consider the coherent regime $M>1$ ($\xi>\xi^{}_{2}$), distinguishing between the adiabatic ($\xi\ll 1$) and non-adiabatic ($\xi\gg 1$) limits. In the coherent and adiabatic limit $\xi^{}_{2}<\xi\ll 1$, we expand $a$ and $b$ in $1/M$ and $P$ to obtain
\begin{align}
\label{eq:34}
&\gamma^{}_{\mathrm{abs}}\approx\frac{\Gamma^{}_{1}}{2}\frac{1-\cos(\tilde{\phi}^{}_{1}-\tilde{\phi}^{}_{2})+f^{}_{1}(\tilde{\phi}^{}_{1},\tilde{\phi}^{}_{2})P}{1-\cos(\tilde{\phi}^{}_{1}-\tilde{\phi}^{}_{2})+f^{}_{1}(\tilde{\phi}^{}_{1},\tilde{\phi}^{}_{2})P+f^{}_{2}(\tilde{\phi}^{}_{1},\tilde{\phi}^{}_{2})(1/M)^{2}},
\end{align}
where $f^{}_{1}(\tilde{\phi}^{}_{1},\tilde{\phi}^{}_{2})=2\cos(\tilde{\phi}^{}_{1}-\tilde{\phi}^{}_{2})-\cos\tilde{\phi}^{}_{1}-\cos\tilde{\phi}^{}_{2}$ and $f^{}_{2}(\tilde{\phi}^{}_{1},\tilde{\phi}^{}_{2})=1/6-(1/24)\cos(\tilde{\phi}^{}_{1}-\tilde{\phi}^{}_{2})$. Hence, $\gamma^{}_{\mathrm{abs}}\approx\Gamma^{}_{1}/2$ with weak dependence on the phases $\tilde{\phi}^{}_{1}$ and $\tilde{\phi}^{}_{2}$. This gives rise to loss tangent $\tan\delta\propto\tan\delta^{}_{0}\cdot\xi^{}_{2}$. Physically, in this regime the probability of photon absorption\textbackslash emission in each transition is $1-P\approx 1$, so that photons are absorbed and reemitted by the TLS, thus dissipated at a rate $\propto\Gamma^{}_{1}$.  

\begin{figure}[htb!]
	\includegraphics[width=\columnwidth,height=7.2cm]{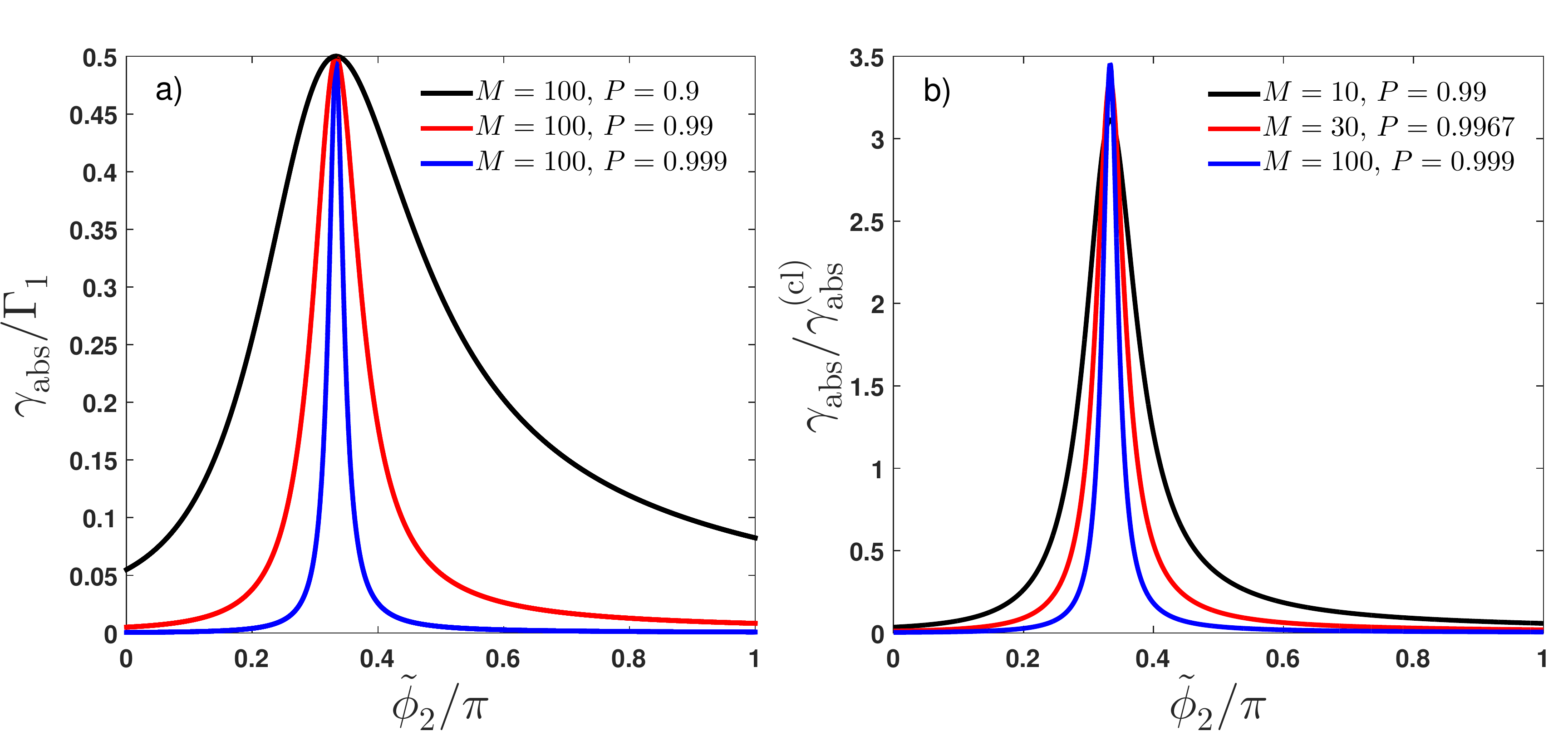}
	\caption{Photon absorption rate per TLS in the non-adiabatic ($\xi\gg 1$ or $P=e^{-1/\xi}\approx 1-1/\xi$) and coherent ($\xi>\xi^{}_{2}$ or $M=\xi/\xi^{}_{2}>1$) regime with $M^{2}(1-P)>1$ (or $\xi>\xi^{2}_{2}$) as a function of the phase $\tilde{\phi}^{}_{2}$ with $\tilde{\phi}^{}_{1}=-\pi/3$. a) In units of the TLS relaxation rate $\Gamma^{}_{1}$ for $M=100$ and various values of $P$ corresponding to the regime $M^{2}(1-P)>1$. In this case the resonance at $\tilde{\phi}^{}_{1}+\tilde{\phi}^{}_{2}=0$ has a constant peak value of $\Gamma^{}_{1}/2$ and width $\delta\phi\propto\sqrt{1-P}$. b) In units of the classical absorption rate $\gamma^{(\mathrm{cl})}_{\mathrm{abs}}$ [Eq.~(\ref{eq:39})] for $\xi^{}_{2}=10$ and $\xi=100$, 300 and 1000, respectively (note that $M=\xi/\xi^{}_{2}$ and $P=e^{-1/\xi}\approx 1-1/\xi$). The resonance at $\tilde{\phi}^{}_{1}+\tilde{\phi}^{}_{2}=0$ corresponds to constructive interference giving rise to rates larger than the classical prediction. The peak of the resonance changes weakly with $\xi$, whereas the width $\propto\sqrt{1-P}\approx1/\sqrt{\xi}$ decreases, leading to a decrease in the resonator loss.} 
	\label{fig2}
\end{figure}

\begin{figure}[htb!]
	\includegraphics[width=\columnwidth,height=7.2cm]{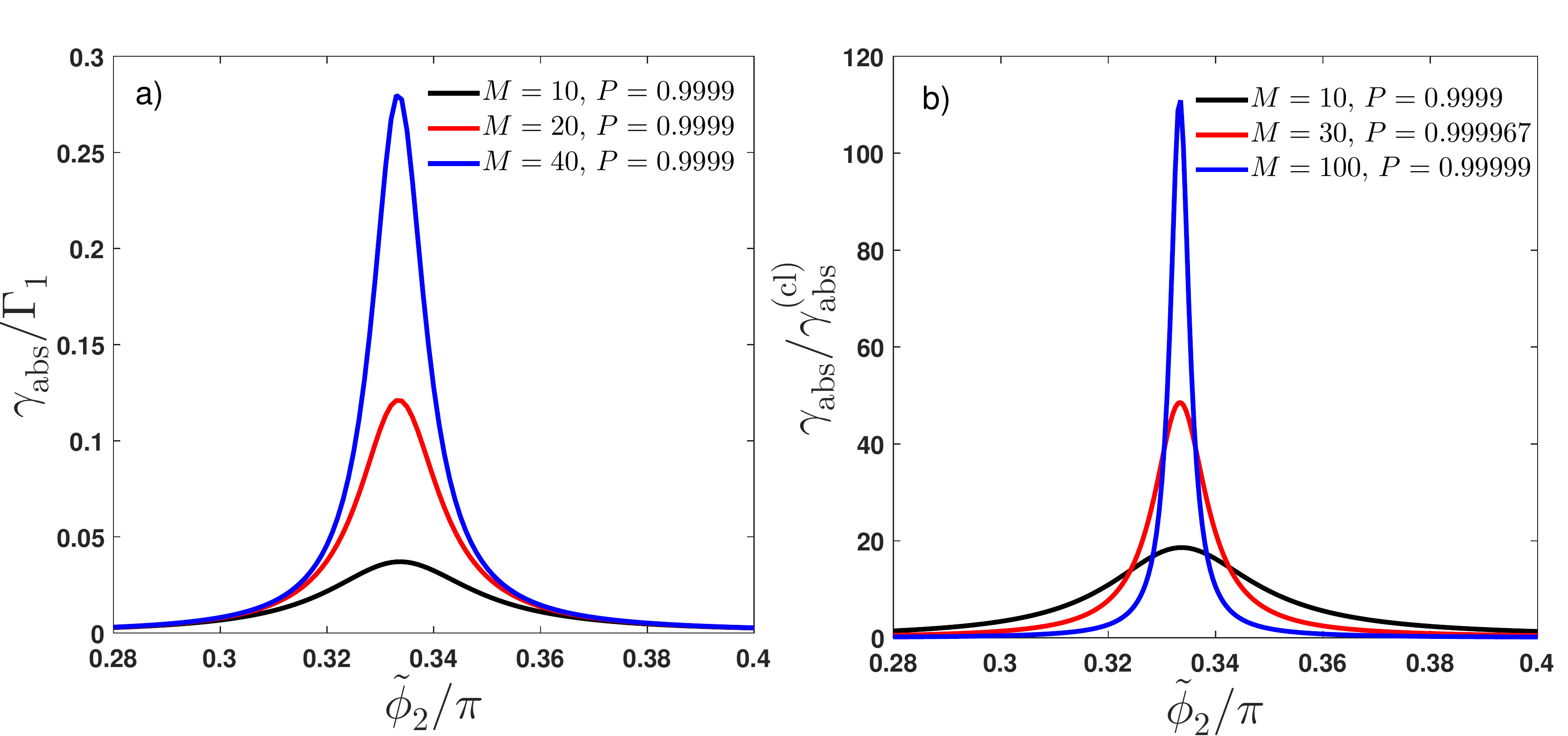}
	\caption{Photon absorption rate per TLS in the non-adiabatic ($\xi\gg 1$ or $P=e^{-1/\xi}\approx 1-1/\xi$) and coherent ($\xi>\xi^{}_{2}$ or $M=\xi/\xi^{}_{2}>1$) regime with $M^{2}(1-P)<1$ (or $\xi^{}_{2}<\xi<\xi^{2}_{2}$) as a function of the phase $\tilde{\phi}^{}_{2}$ with $\tilde{\phi}^{}_{1}=-\pi/3$. a) In units of the TLS relaxation rate $\Gamma^{}_{1}$ for $P=0.9999$ and various values of $M$ corresponding to the regime $M^{2}(1-P)<1$. In this case the peak of the resonance at $\tilde{\phi}^{}_{1}+\tilde{\phi}^{}_{2}=0$ scales as $M^{2}(1-P)$ and the width as $\delta\phi\propto 1/M$. Note that the contribution of this resonance to the absorption rate is $\gamma^{\mathrm{res}}_{\mathrm{abs}}\approx\gamma^{}_{\mathrm{abs}}(\tilde{\phi}^{}_{1}=-\tilde{\phi}^{}_{2})\cdot\delta\phi\propto\Gamma^{}_{1}/\xi^{}_{2}$, and therefore the loss tangent $\tan\delta\propto\xi^{}_{2}\cdot\gamma^{\mathrm{res}}_{\mathrm{abs}}/\Gamma^{}_{1}$ does not increase with $M$ in this regime. b) In units of the classical absorption rate $\gamma^{(\mathrm{cl})}_{\mathrm{abs}}$ [Eq.~(\ref{eq:39})] for $\xi^{}_{2}=1000$ and $\xi=10^{4}$, $3\cdot 10^{4}$ and $10^{5}$, respectively. The resonance at $\tilde{\phi}^{}_{1}+\tilde{\phi}^{}_{2}=0$ corresponds to constructive interference giving rise to rates larger than the classical prediction. The peak of the resonance grows with $\xi$, whereas the width $\propto1/M\propto 1/\xi$ decreases, leading to a weak dependence of the resonator loss on $\xi$.}
	\label{fig3}
\end{figure}

In the coherent and non-adiabatic limit $\xi\gg 1,\xi^{}_{2}$ we expand $a$ and $b$ in $1/M$ and $1-P$ to obtain
\begin{align}
\label{eq:35}
&\gamma^{}_{\mathrm{abs}}\approx\frac{\Gamma^{}_{1}}{2}\frac{(1-P)\left(2-\cos\tilde{\phi}^{}_{1}-\cos\tilde{\phi}^{}_{2}\right)}{1-\cos(\tilde{\phi}^{}_{1}+\tilde{\phi}^{}_{2})+g^{}_{1}(\tilde{\phi}^{}_{1},\tilde{\phi}^{}_{2})(1-P)+g^{}_{2}(\tilde{\phi}^{}_{1},\tilde{\phi}^{}_{2})(1/M)^{2}},
\end{align}
where $g^{}_{1}(\tilde{\phi}^{}_{1},\tilde{\phi}^{}_{2})=2\cos(\tilde{\phi}^{}_{1}+\tilde{\phi}^{}_{2})-\cos\tilde{\phi}^{}_{1}-\cos\tilde{\phi}^{}_{2}$ and $g^{}_{2}(\tilde{\phi}^{}_{1},\tilde{\phi}^{}_{2})=2/3-(13/24)\cos(\tilde{\phi}^{}_{1}+\tilde{\phi}^{}_{2})$. We observe that a resonance occurs at $\tilde{\phi}^{}_{1}+\tilde{\phi}^{}_{2}=\phi^{}_{1}+\phi^{}_{2}=2\pi n$, where $n$ is an integer. Note that Eq.~(\ref{eq:32}) is a $2\pi$-periodic function of $\phi^{}_{1}$ and $\phi^{}_{2}$, and it is therefore sufficient to consider the range $\phi^{}_{1},\phi^{}_{2}\in[-\pi,\pi)$, where only the resonance with $n=0$ is relevant. This resonance is shown in Figs.~\ref{fig2} and~\ref{fig3}. Outside the resonance, the last two terms in the denominator of Eq.~(\ref{eq:35}) are negligible and $\gamma^{\mathrm{non-res}}_{\mathrm{abs}}\propto\Gamma^{}_{1}(1-P)\approx\Gamma^{}_{1}/\xi$, with weak dependence on the phases $\tilde{\phi}^{}_{1}$ and $\tilde{\phi}^{}_{2}$. Let us evaluate the contribution of the resonance. According to Eq.~(\ref{eq:35}), the resonance width is 
$\delta\phi\propto\max\{\sqrt{1-P},1/M\}$, i.e.\ $\delta\phi\propto\sqrt{1-P}$ for $M^{2}(1-P)>1$ ($\xi>\xi^{2}_{2}$) and $\delta\phi\propto 1/M$ for $M^{2}(1-P)<1$ ($\xi<\xi^{2}_{2}$). The peak value of the resonance is constant and equals $\Gamma^{}_{1}/2$ for $M^{2}(1-P)>1$, whereas it scales as $M^{2}(1-P)$ for $M^{2}(1-P)<1$. These two different behaviors of the resonance are shown in Fig.~\ref{fig2}a) and Fig.~\ref{fig3}a), respectively. Based on the random ensemble of TLSs, it is plausible to assume that the phases $\phi^{}_{1}$ and $\phi^{}_{2}$ are distributed homogeneously, without special preference to the resonance condition $\phi^{}_{1}+\phi^{}_{2}=2\pi n$. The contribution of this resonance to the averaged absorption rate can therefore be estimated as
\begin{align}
\label{eq:36}
&\gamma^{\mathrm{res}}_{\mathrm{abs}}\approx\gamma^{}_{\mathrm{abs}}(\tilde{\phi}^{}_{1}=-\tilde{\phi}^{}_{2})\cdot\delta\phi\propto\Gamma^{}_{1}\cdot\begin{cases}
\sqrt{1-P}\approx 1/\sqrt{\xi} & M^{2}(1-P)>1 \\ 
M(1-P)\approx 1/\xi^{}_{2} & M^{2}(1-P)<1
\end{cases}.
\end{align}
Note that in both cases $\gamma^{\mathrm{res}}_{\mathrm{abs}}>\gamma^{\mathrm{non-res}}_{\mathrm{abs}}$, so that the contribution of the resonance is the dominant one.

\begin{figure}[htb!]
	\includegraphics[width=0.7\textwidth,height=7.5cm]{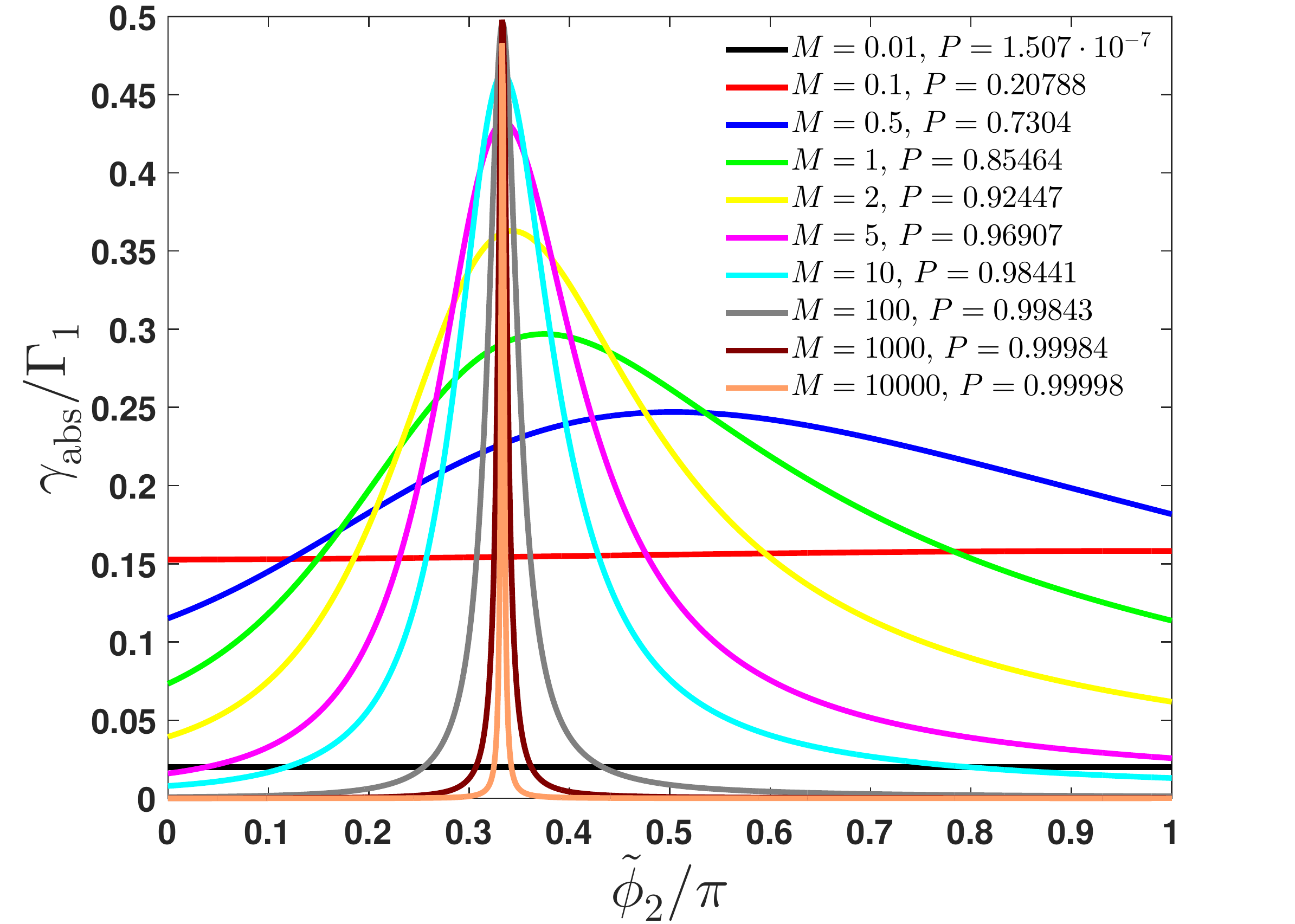}
	\caption{Photon absorption rate per TLS (in units of the TLS relaxation rate $\Gamma^{}_{1}$) as a function of the phase $\tilde{\phi}^{}_{2}$ with $\tilde{\phi}^{}_{1}=-\pi/3$, for $\xi^{}_{2}=10$ and $\xi=0.1$, 1, 5, 10, 20, 50, $100$, $10^{3}$ and $10^{4}$, respectively (the legend shows the values of $M=\xi/\xi^{}_{2}$ and $P=e^{-1/\xi}$).}
	\label{fig4}
\end{figure}

The behavior of the absorption rate per TLS is summarized in Fig.~\ref{fig4}, which plots $\gamma^{}_{\mathrm{abs}}$ as a function of one of the phases (similarly to Figs.~\ref{fig2} and~\ref{fig3}) for various values of $\xi$ for a fixed $\xi^{}_{2}=10$. For $M<1$ ($\xi<\xi^{}_{2}$) the absorption rate weakly depends on the phases, corresponding to value given by Eq.~(\ref{eq:29}). The resonance at $\phi^{}_{1}+\phi^{}_{2}=2\pi n$ develops at $M\approx 1$. For $M^{2}(1-P)<1$ ($\xi^{}_{2}<\xi<\xi^{2}_{2}$) the area of the resonance depends weakly on $\xi$, whereas it decreases with $\xi$ for $M^{2}(1-P)>1$ ($\xi>\xi^{2}_{2}$) since its peak is constant and equals $\Gamma^{}_{1}/2$ and its width decreases, in accordance with Eq.(\ref{eq:36}).

\begin{figure}[htb!]
	\includegraphics[width=0.7\textwidth,height=7.5cm]{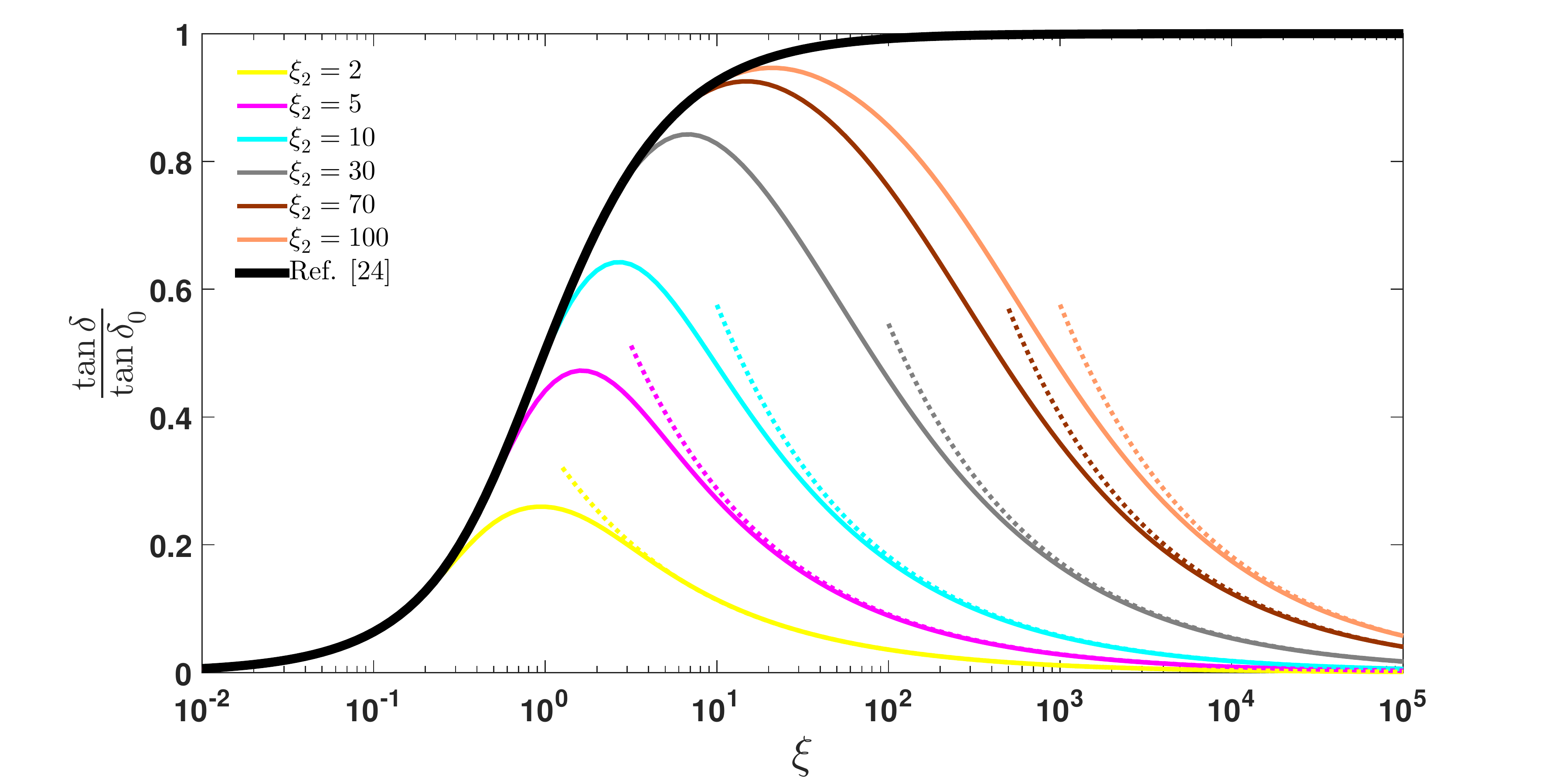}
	\caption{Theoretical results for the loss tangent due to TLSs, normalized by the intrinsic low-power loss tangent $\tan\delta^{}_{0}=\pi P^{}_{0}p^{2}/(3\epsilon)$, as a function of the dimensionless sweep rate $\xi\equiv 2|v^{}_{0}|/(\pi\hbar\Omega^{2}_{\mathrm{R}0})$ for various values of $\xi^{}_{2}\equiv 8pE^{}_{\mathrm{max}}\Gamma^{}_{1}/(\pi\hbar\Omega^{2}_{\mathrm{R}0})$ in the regime $\xi^{}_{2}>1$ (same results as in Fig.~2 in the main text). Dotted curves are fits to the form $A\xi^{}_{2}/\sqrt{\xi}$ with a single fit parameter $A=0.18$. This form of the asymptotic decrease is valid for $\xi>\xi^{2}_{2}$.} 
	\label{fig5}
\end{figure}

The above analysis reveals the following qualitative behavior in the coherent regime $\xi>\xi^{}_{2}$. For $\xi^{}_{2}<1$ the loss tangent behaves as
\begin{align}
\label{eq:37}
\frac{\tan\delta}{\tan\delta^{}_{0}}\propto\begin{cases}
\xi & \xi<\xi^{}_{2} \\ 
\xi^{}_{2} & \xi^{}_{2}<\xi\ll 1 \\
\xi^{}_{2}/\sqrt{\xi} & \xi\gg 1
\end{cases},
\end{align} 
whereas for $\xi^{}_{2}>1$ 
\begin{align}
\label{eq:38}
\frac{\tan\delta}{\tan\delta^{}_{0}}\propto\begin{cases}
\xi & \xi\ll 1 \\ 
1 & 1\ll\xi<\xi^{}_{2} \\
1 & \xi^{}_{2}<\xi\ll\xi^{2}_{2} \\
\xi^{}_{2}/\sqrt{\xi} & \xi\gg\xi^{2}_{2}
\end{cases}.
\end{align}
Note that for $\xi^{}_{2}>1$ the scale that governs the decrease in loss tangent as $1/\sqrt{\xi}$ is $\propto\xi^{2}_{2}$. This is demonstrated in Fig.~\ref{fig5}, showing some of the curves from Fig.~2 in the main text corresponding to the regime $\xi^{}_{2}>1$. The asymptotic decrease in loss is fitted to $\tan\delta/\tan\delta^{}_{0}=A\xi^{}_{2}/\sqrt{\xi}$ using a single fit parameter $A$ for all curves. As one can observe, this functional form describes the data very well at the regime $\xi>\xi^{2}_{2}$. 

One can better understand the importance of interference between multiple LZ transitions for the reduction in the resonator loss at the coherent and non-adiabatic regime ($\xi\gg 1,\xi^{}_{2}$) by comparing the above results with those obtained from a classical approach based on rate equations, as described in the main text [Eqs.~(11) and~(12) in the main text]. Using this classical approach, the photon absorption rate per TLS is found to be
\begin{align}
\label{eq:39}
\gamma^{(\mathrm{cl})}_{\mathrm{abs}}\approx\gamma/(1+2\gamma/\Gamma^{}_{1}),
\end{align} 
where $\gamma=2(1-P)/T^{}_{\mathrm{sw}}$ is the photon emission and absorption rate in a single transition [see Eq.~(\ref{eq:29})]. This expression reduces to $\gamma^{(\mathrm{cl})}_{\mathrm{abs}}\approx\gamma$ for $\gamma\ll\Gamma^{}_{1}$ (or, equivalently, $M(1-P)\ll 1$) and $\gamma^{(\mathrm{cl})}_{\mathrm{abs}}\approx\Gamma^{}_{1}/2$ for $\gamma\gg\Gamma^{}_{1}$ (or $M(1-P)\gg 1$), which can be summarized as $\gamma^{(\mathrm{cl})}_{\mathrm{abs}}\approx\mathrm{min}\{\gamma,\Gamma^{}_{1}/2\}$. This is clear, since a TLS cannot dissipate photons at a rate faster than its decay rate (due to coupling to its own bath, e.g., phonons). In Figs.~\ref{fig2}b) and~\ref{fig3}b) we compare the absorption rate obtained by the theory described above with the classical prediction of Eq.~(\ref{eq:39}). One sees that the resonant peak at $\tilde{\phi}^{}_{1}+\tilde{\phi}^{}_{2}=\phi^{}_{1}+\phi^{}_{2}=2\pi n$ corresponds to constructive interference, yielding absorption rates larger than the classical result $\gamma^{(\mathrm{cl})}_{\mathrm{abs}}$. The region outside this resonance corresponds to destructive interference with rates smaller than $\gamma^{(\mathrm{cl})}_{\mathrm{abs}}$. As shown in Fig.~\ref{fig2}b), in the regime $M^{2}(1-p)>1$ ($\xi>\xi^{2}_{2}$) the peak value depends weakly on $\xi$, but its width decreases as $\sqrt{1-P}\approx1/\sqrt{\xi}$, leading to the reduction of the loss tangent as $1/\sqrt{\xi}$. On the other hand, in the regime $M^{2}(1-p)<1$ ($\xi^{}_{2}<\xi<\xi^{2}_{2}$) the enhancement in the peak value roughly cancels the decrease in its width, leading to weak dependence of the loss tangent on $\xi$. The classical approach cannot capture the reduction in loss at high sweep rates, but instead predicts the loss tangent to be  
\begin{align}
\label{eq:40}
\frac{\tan\delta}{\tan\delta^{}_{0}}\propto\begin{cases}
\xi & \xi<\xi^{}_{2} \\ 
\xi^{}_{2} & \xi^{}_{2}<\xi
\end{cases}
\end{align} 
for $\xi^{}_{2}<1$, and 
\begin{align}
\label{eq:41}
\frac{\tan\delta}{\tan\delta^{}_{0}}\propto\begin{cases}
\xi & \xi\ll 1 \\ 
1 & 1\ll\xi
\end{cases}
\end{align}
for $\xi^{}_{2}>1$. The decrease in the resonator loss is therefore attributed to interference between multiple transitions.
\section{Experimental details} \label{Exp}

\subsection{Setup}
\begin{figure}
	\includegraphics[width=12cm,height=8cm]{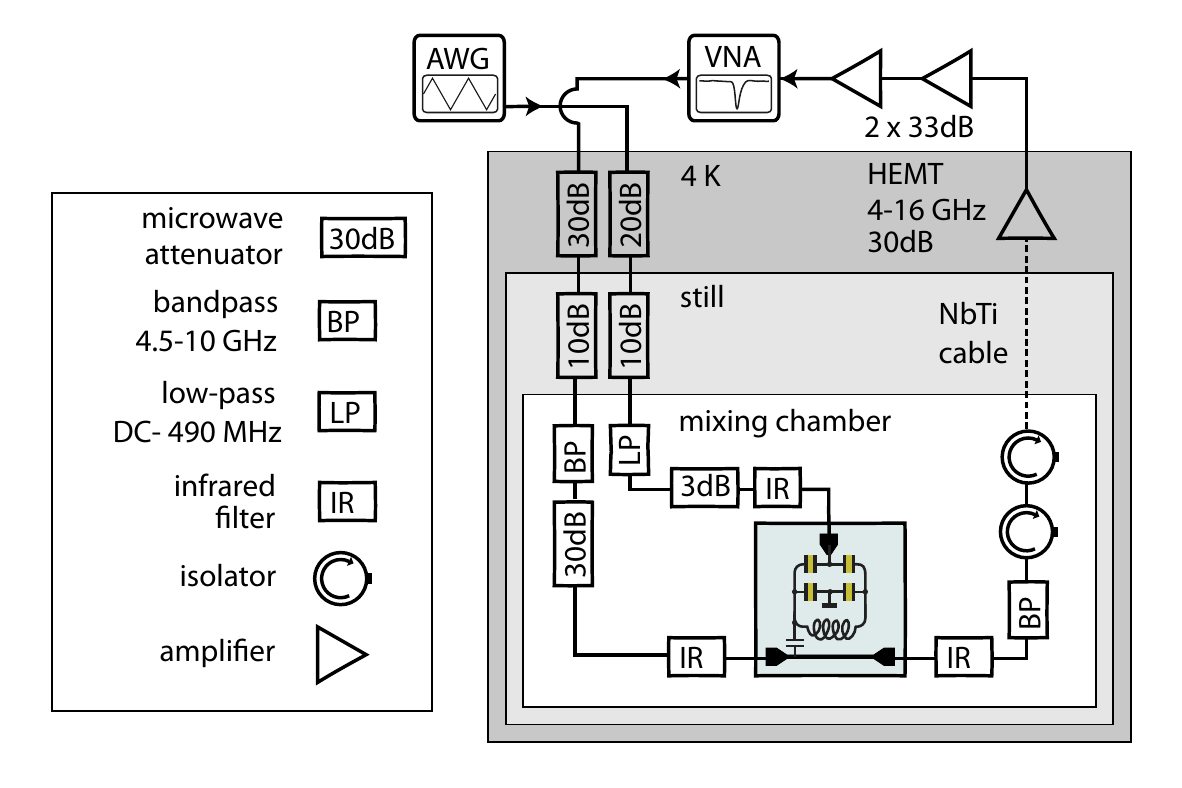}\protect\caption{\label{fig:supp_setup}
	Setup schematic. Resonator loss is measured by a Vector Network Analyzer (VNA). Thermal isolation of coax cables and noise protection is provided by attenuators thermally anchored to different temperature stages, band-pass (BP) and low-pass filters (LP) (Mini-Circuits), and custom-made infrared filters (IR). After interacting with the sample, the signal passes through two magnetically shielded isolators (QuinStar QCY-060400CM00), a high-mobility electron transistor (HEMT) (Low-Noise Factory LNF-LNC4-16A) and two room-temperature amplifiers (Mini-Circuits VA-183-S). The electric bias to the resonator dielectric is provided by an arbitrary waveform generator (Tektronix AWG 5914B, 1.2 GS/s), generating symmetric triangular waveforms of minimal duration 10ns and 4V amplitude.}
\end{figure}

The sample was measured in an Oxford Kelvinox 100 dilution refrigerator at a temperature of 30 mK. The sample chip is installed a a light-tight aluminum housing surrounded by a cryoperm magnetic shield. To enable measurements of the resonator in the single-photon regime, the microwave circuitry is heavily filtered and attenuatored as illustrated in Fig.~\ref{fig:supp_setup}. 

\subsection{Sample fabrication} \label{app:samplefab}
Samples are fabricated in three optical lithography steps. First, a $50\,$nm thick layer of aluminum is deposited at a rate of $1\,$nm/s on a cleaned sapphire substrate using an electron beam evaporation system (Plassys MEB 550 S). This layer is patterned into ground plane, transmission line, and resonator inductances using S1805 photoresist, AZ-developer, and reactive-ion etching in a ICP machine (Oxford 100 ICP 180), before stripping the remaining photoresist with NEP. We preferred to etch the first layer over a lift-off process to avoid contamination of the substrate-metal interface by photoresist residuals, while further lithography steps are done by a lift-off process to avoid the first layer to be damaged by etching. AZ developer was used because we found that it produces less defects in underlying aluminum films compared to MF319 developer.\\

In the second lithography step, prior to deposition of the capacitor dielectric, the native aluminum oxide on the bottom capacitor electrodes is removed using an argon ion mill~\cite{GL17}. The aluminum oxide is then deposited at a rate of $0.3\,$nm/s by aluminum evaporation while the vacuum chamber is exposed to an oxygen flow of 5~sccm, and afterwards covered with a $30\,$nm thick top layer of aluminum. In a third lithography step, after initial ion-milling removal of the native aluminum oxide, the vias which electrically connect bottom and top metallic layers [see Fig. 3a) in the main text] are formed by placing a $50\,$nm-thick layer of aluminum as a bandage overlapping both layers. \\

\subsection{Sample parameters}
Table~\ref{tab:resparms} describes the sample parameters of three different resonators.
\begin{table}
	\begin{tabular}{cccccccc}
		\hline\hline
		Resonator & $f^{}_{\mathrm{res}}$ ($\mathrm{GHz}$) & $w^{}_{\mathrm{cap}}$ ($\mu\mathrm{m}$) & $d^{}_{\mathrm{coup}}$ ($\mu\mathrm{m}$) & $Q^{}_{\mathrm{c}}\ (10^3)$  & $\tan\delta^{}_{\mathrm{b}}\ (10^{-4})$ & & $\tan\delta^{}_{0}\ (10^{-4})$  \\\hline
		1 & 6.892 & 5 & 100 & 12.8\,$\pm\,$0.7 & $8.85$ & & $3.03$ \\
		2 & 6.952 & 7 & 18 & 3.4\,$\pm\,$0.3 & $17$ & & $3.4$ \\
		\hline\hline
	\end{tabular}
	\caption{Resonators parameters: measured resonance frequency $f^{}_\mathrm{fes}$, lateral size of square capacitors $w^{}_{\mathrm{cap}}$, coupling distance to transmission line $d^{}_{\mathrm{coup}}$, coupling quality factor $Q^{}_{\mathrm{c}}$, background loss tangent $\tan\delta^{}_{\mathrm{b}}$, and intrinsic TLS loss tangent $\tan\delta^{}_{0}$}.
	\label{tab:resparms}
\end{table}

\section{Additional data} \label{data}
\begin{figure}
\includegraphics[width=0.8\textwidth]{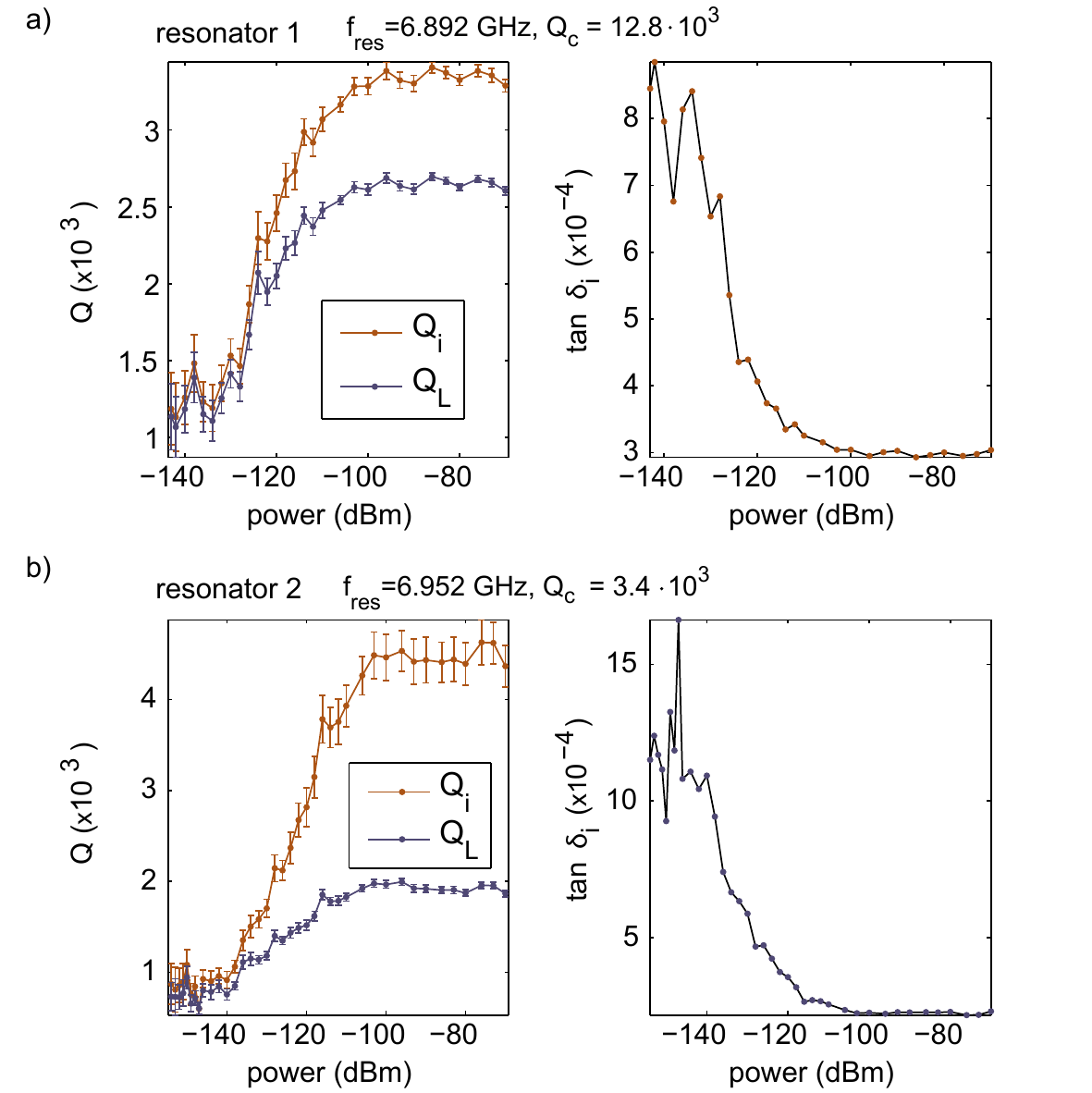}\protect\caption{\label{fig:supp5}
a) Internal quality factor $Q^{}_{i}$ and measured loaded quality factor $Q^{}_{L}$ (left panel), and resonator loss tangent $\tan\delta^{}_{i}\equiv 1/Q^{}_{i}$ (right panel) vs.\ power at the input of the resonator, for resonator 1. Its resonance frequency is $f^{}_{\mathrm{res}}=\omega/(2\pi)=6.892\,$GHz and the coupling quality factor is $Q^{}_{c}=12.3\cdot 10^{3}$.
b) Same measurements for resonator 2, having a resonance frequency of $f^{}_{\mathrm{res}}=\omega/(2\pi)=6.952\,$GHz and a coupling quality factor of $Q^{}_{c}=3.4\cdot 10^{3}$.}
\end{figure}

Figure~\ref{fig:supp5} shows the resonator internal quality factor $Q^{}_{i}$ and the measured loaded quality factor $Q^{}_{L}$ as a function of input power, in the absence of a bias field. These curves show the saturation of the TLS photon absorption with increasing power.

